\documentclass[useAMS,usenatbib,usegraphicx,onecolumn]{mn2e}

\def\lapp{\ \lower 3pt\hbox{${\buildrel < \over \sim}$}\ }
\def\gapp{\ \lower 3pt\hbox{${\buildrel > \over \sim}$}\ }

\newcommand{\be}{\begin{equation}}
\newcommand{\ee}{\end{equation}}
\newcommand{\bea}{\begin{eqnarray}}
\newcommand{\eea}{\end{eqnarray}}
\newcommand{\next}{\nonumber\\}
\newcommand{\rn}[1]{(\ref{#1})}
\newcommand{\ff}[2]{{\textstyle \frac{#1}{#2}}}
\newcommand{\hand}{\hspace{0.5cm}{\rm and}\hspace{0.5cm}}
\newcommand{\oW}{\overline{W}_q}
\newcommand{\ms}{{\rm m\, s}^{-1}}

\newcommand{\ben}{\begin{enumerate}}
\newcommand{\een}{\end{enumerate}}

\title[Companions to short-period planets]
{Long-term tidal evolution of short-period planets with companions}
\author[Rosemary A. Mardling]{Rosemary A. Mardling$^{1}$\thanks{E-mail:
mardling@sci.monash.edu.au}\\
$^{1}$School of Mathematical Sciences, Monash University, Victoria, 3800, Australia}
\begin{document}

\date{Accepted ... Received ...; in original form ...}

\pagerange{\pageref{firstpage}--\pageref{lastpage}} \pubyear{2007}

\maketitle

\label{firstpage}

\begin{abstract}

Of the fourteen transiting extrasolar  
planetary systems for which radii have been measured, 
at least three appear to be
considerably larger than theoretical estimates suggest.
It has been proposed by Bodenheimer, Lin \& Mardling that
undetected companions acting to excite the orbital eccentricity
are responsible for these oversized planets, 
as they find new equilibrium radii in response to being tidally heated.
In the case of HD 209458, this hypothesis has been rejected by
some authors because
there is no sign of such a companion at the 5 ms$^{-1}$ level,
and because it is difficult to say conclusively that the eccentricity is non-zero.
Transit timing analysis as well
as a direct transit search have further constrained the existence of very short-period
companions, especially in resonant orbits.
Whether or not a companion is responsible for the large
radius of HD 209458b, almost certainly some short-period systems
have companions which force their eccentricities to
nonzero values. This paper is dedicated to quantifying
this effect.

The eccentricity of a short-period planet will
only be excited as long as its (non-resonant) companion's eccentricity is non-zero.
Here we show that the latter decays on a timescale which
depends on the structure of the {\it interior} planet,
a timescale which is often shorter than the lifetime of the system.
{\it This includes Earth-mass planets in the habitable zones of some stars}.
We determine which configurations are capable of sustaining 
significant eccentricity for at least the age of the system,
and show that these include systems with companion masses
as low as a fraction of an Earth mass.
The orbital parameters of such companions are consistent with
recent calculations
which show that the migration process can induce the formation
of low mass planets external to the orbits of hot Jupiters.
Systems with inflated planets are therefore good targets
in the search for terrestrial planets.

\end{abstract}


\section{Introduction}

While most extrasolar planet 
discoveries to date have relied on the radial velocity method
of detection with its inherent uncertainty in the planetary mass,
a few systems have been observed in
transit, thereby breaking this degeneracy. Not only does this allow a
reasonably accurate determination of the planetary mass (given a
reliable estimate for the stellar mass is available), it allows one to 
estimate the radius of the planet, and hence its density. Before the
discovery that HD 209458b transited its host star \citep{c2,h1}, 
it was by no means clear that the
planets discovered till then, all with masses similar to or more than
Jupiter's, were gas giants. The extra information gleaned from
photometric transit data has provided the opportunity to study planetary
{\it structure} \citep{b4},
while spectroscopic data have revealed their chemical composition \citep{c5}.
In the former case, determination of the planetary radius allows one to
put constraints on the possible existence of a planetary core, and this in turn
allows one to constrain the mode of planet formation, at least for that
planet. In principle this allows a choice 
between the core accretion model \citep{s1,w1,b5}
and the gravitational instability model \citep{k2,c3,b6}, although it is not
clear that the latter should produce core-less planets given 
accretion of solids plus subsequent gravitational settling 
almost certainly follows formation. Either way,
such an exercise requires knowledge of the equilibrium structure of 
a gas giant planet, with and without a core, and with and without external and
internal heating. 

While it can be argued that there are problems matching theoretical radii with
observation for {\it all} transiting planets \citep{b2},
three stand out as particularly
excessive. These are HD 209458b \citep{c6,h1,k1}, WASP-1b \citep{c1} and
HAT-P-1b \citep{b1}, for which the planetary radii and other observationally
derived parameters are
listed in Table~\ref{planets}. Data for HD 209458 and HAT-P-1 are taken from 
Burrows et al.\ (2006) and references therein, except for the eccentricities,
while data for WASP-1 are taken from \citet{s2}.
\begin{table*}
\label{planets}
 \centering
 \begin{minipage}{140mm}
  \caption{Observational data for inflated planets}
\begin{tabular}{ccccccc}
\hline
Planet & Period (day) & $m_*$ ($M_\odot$) & $m_p$ ($M_J$) & $a_p$ (AU) & $e_p^{obs}$ & $R_p$ ($R_J$) \\
\hline
HD 209458b & 3.5247 & 1.10 & $0.64\pm0.06$ & 0.045 & 
$0.014\pm0.009$\footnote{\citet{l1}.} & $1.32\pm0.03$ \\
WASP-1b\footnote{\citet{s2}.} & 2.5199 & 1.15 & $0.79^{+0.13}_{-0.06}$ & 0.0379 & - &  $1.443\pm0.039$ \\
HAT-P-1b & 4.4653 & 1.12 & $0.53\pm0.04$ & 0.0551 &$0.09\pm0.02$\footnote{This
value is based on very little data \citep{b1}.}& $1.36^{+0.11}_{-0.09}$
\\
\hline
\end{tabular}
\end{minipage}
\end{table*}
Here $m_*$ and $m_p$ are the masses of the star and planet respectively,
$a_p$ and $e_p$ are the orbital semimajor axis and eccentricity
and $R_p$ is
the estimated planetary radius ($R_J$ is Jupiter's radius).
The eccentricity of HD 209458b is from \citet{l1}, a paper devoted
specifically to discussing this contentious orbital element,
while that of HAT-P-1b is from \citet{b1}, a value based on 
very little data. However, as pointed out by the latter authors, HAT-P-1b is
further from its parent star than HD 209458b, and is such that
the relatively large estimated value of 0.09 produces a similar
planetary radius.

Various mechanisms have been suggested to account for the excess
radii of the planets listed in Table~\ref{planets}. Obvious contenders
are stellar irradiation \citep{b7,g2,b3,b8,a3,f4,b2} and tidal forcing by	
a companion planet \citep{b4,b3}, the latter requiring non-zero eccentricity
of the observed planetary orbit to be viable. 

A less obvious mechanism has been proposed by \citet{w4} in which
``obliquity tides'' are forced if the system can manage to remain
in a suitable Cassini state. 
Both \citet{l4} and \citet{f3} verify that in principle it is possible for this mechanism to operate as long as the
system dissipates energy at a rate which is too small to inflate the planet to 
the observed size.
 
This paper focuses on the mechanism of tidal forcing provided by a
companion planet, and hence requires that the orbit of the 
transiting planet be non-circular.
Using a bootstrap Monte Carlo technique
to estimate confidence intervals for
the eccentricity of HD 209458b, \citet{l1} 
find that $e_p<0.049$ with 99.73\% confidence, 
$e_p<0.034$ with 95.4\% confidence, and
$e_p<0.02$ with 68.3\% confidence,
with longitude of periastron
$\varpi_p=80^o\pm 80^o$. Their best-fit measurement 
for the eccentricity is $e_p=0.014$.
However, they point out that because $e_p$ has a hard lower bound
of zero, any best-fit model to a radial velocity data set will have $e_p\ge 0$,
with largest estimates corresponding to $\varpi_p\simeq 90^o$ and
$270^o$. Using synthetic data 
corresponding to a truly circular orbit, such bootstrap trials produce
a distribution in $\varpi_p-e_p$ space which is symmetric about these two values,
while use of the real radial velocity data produces a distinctly asymmetric
distribution which is biased towards $\varpi_p=90^o$. 
This suggests that in fact $e_p\neq 0$, although $e_p=0$ cannot be ruled out.
\citet{l1} point out that
the fact that the estimate for $\varpi_p$ is centred near $90^o$ suggests
that even if HD 209458b has a non-zero eccentricity, timing of the
secondary eclipse will find $e_p\cos\varpi_p\simeq 0$, as has recently been found
by \citet{d1} in the infrared. Note also that the orbital model of \cite{l1} leads
to a stellar jitter estimate of 3.3 $\ms$ which is close to the predicted
value of 2.8 $\ms$. Thus, at least for HD 209458, we concentrate our attention 
in this paper on
companions which produce stellar reflex velocities not much larger than this value.

The presence of a very short-period companion to HD 209458b has
been ruled out by transit-timing analysis for interior and exterior
planets with periods up to 21 days (corresponding to the 6:1 resonance
at 0.16 AU) \citep{a2},
as well as by a direct transit search for planets with radii down
to $2 R_\oplus$ and periods up to 14 days
\citep{c4}. Here we will concentrate
on companion planets with semimajor axes greater than these lower bounds.

This paper
presents a generic analysis for short-period systems with external companion
planets (low-mass or otherwise), estimating
the range of orbital parameters
and masses required for such companions to be responsible for
inflated radii. 
An unexpected result is that very low mass planets
are sometimes capable of inducing significant tidal forcing on
extremely long timescales.
Section 2 discusses the relationship between the eccentricity
and the $Q$-value of a planet
for a given value of the planet's luminosity,
the $Q$-value or  tidal quality factor \citep{g1}
being a measure of how efficiently the tides dissipate their oscillation energy.
Section 3 discusses three-body dynamics in the presence of dissipation
(see also \citet{m3} and \citet{a0}).
We start by 
reviewing the secular evolution of point-mass coplanar three-body systems,
with and without relativistic effects, 
deriving explicit formulae for the periods of variation of the eccentricities
and their amplitudes. We identify the orbital parameters which
correspond to libration and circulation of 
the angle $\eta\equiv\varpi_p-\varpi_c$, where
$\varpi_c$ is the longitude of periastron of the
companion, and in particular find
the fixed points in $e_p-\eta$ space.
We then show that tidal damping
results in the system evolving towards the relevant fixed point
(now a {\it quasi}-fixed point)
on three times the tidal damping timescale. Once this is achieved the system
continues to evolve toward the true fixed point for which both
eccentricities are zero. The timescale for this latter phase may be
several orders of magnitude 
longer than the tidal damping timescale, and can in fact be much longer
than the age of the system. We illustrate the theory for
hypothetical companions to HD 209458,
and show that companion
masses as low as a fraction of an Earth mass are capable of 
sustaining non-zero eccentricites in the observed planetary orbit for at 
least the age of the system, and in some cases at the level required
to sustain the observed radii at the current epoch.

Section 4 maps out regions in $m_c-a_c$ parameter space,
where $m_c$ and $a_c$ are the mass and semimajor axis
of a hypothetical companion, for which the eccentricities of HD 209458b and
HAT-P1b would be at their observed level, as well as for
WASP-1b which for which the eccentricity is currently unconstrained.
Section 5 applies the secular theory developed here to determine which
configurations have circularized companion planets, while
Section 6 presents a summary and a discussion.

\section{The Eccentricity-Radius Relation}

Various authors have studied the problem of the response of 
a planet to internal and external heat sources (eg.\ Burrows et al.\ 2000,
Bodenheimer, Lin \& Mardling 2001,
Bodenheimer, Laughlin \& Lin 2003, Burrows et al.\ 2007).
In particular, models include planets with and without cores, with and without
stellar irradiation, and with and without an internal  heat source.
In all these models the planet assumes an equilibrium radius, $R_p$,
and it is the relationship between this and the rate of energy dissipation in the planet,
${\cal L}_p$, that is used in an attempt to explain the range of 
planetary radii observed in transiting planets.

In the case that ${\cal L}_p$ is due to tides raised by an external companion,
the rate of internal energy dissipation in the planet is given by
(eg., \citet{y1}, \citet{b3})
\be
\dot{E}_d=\left(\frac{Gm_*m_p}{a_p}\right)\left(\frac{e_p^2}{\tau_{circ}}\right),
\ee
to second order in the eccentricity, where
\be
\tau_{circ}\equiv\frac{e_p}{\dot e_p}
=\frac{2}{21n_p}\left(\frac{Q_p}{k_p}\right)\left(\frac{m_p}{m_*}\right)
\left(\frac{a_p}{R_p}\right)^5
\label{tcirc}
\ee
is the circularization timescale of the orbit, with $n_p$ the orbital
frequency,
$k_p$ the tidal Love number of the planet and
$Q_p$ its $Q$-value or
tidal quality factor (Goldreich \& Soter 1966).\footnote{Note that
Goldreich \& Soter (1966) use a modified $Q$-value, $Q'_p$, which absorbs
the Love number such that $Q_p'=3Q_p/2k_p$; equation (4) in
\citep{b3} should have $Q'_p$ instead of $Q_p$.
Another potential source of
confusion is the use of the (quadrupole)
apsidal motion constant $k_2$ rather than the
tidal Love number; these are related by $k_p=2k_2$.}
Equating this to the power, ${\cal L}_p$, 
needed to maintain an observed planetary radius
gives $e_p$ in terms of the observable quantities $m_*$, $R_p$ and $a_p$ 
($m_p$ actually drops out), as well as the theoretical quantities 
$Q_p$, $k_p$ and ${\cal L}_p$:
\be
e_p=0.020\left(\frac{m_*}{M_\odot}\right)^{-5/4}
\left(\frac{R_p}{R_J}\right)^{-5/2}
\left(\frac{a_p}{0.04{\rm AU}}\right)^{15/4}
\left(\frac{Q_p}{10^5}\right)^{1/2}
\left(\frac{k_p}{k_J}\right)^{-1/2}
\left(\frac{{\cal L}_p}{10^{-8}L_\odot}\right)^{1/2},
\label{ept}
\ee
where $k_J=0.34$ is the tidal Love number for Jupiter.
\citet{b2} calculate ${\cal L}_p$ 
{\it for coreless solar metalicity planets which are irradiated by
their parent stars}, values for which are listed in Table~\ref{theory}
under ${\cal L}_p^{Bu}$.
\begin{table*}
\label{theory}
 \centering
 \begin{minipage}{140mm}
  \caption{Theoretical data assuming $Q_p=10^5$}
\begin{tabular}{ccccccc}
\hline
Planet & Age (Gyr) & $\tau_{circ}$ (Gyr)
& ${\cal L}_p^{Bu}$ ($L_\odot$) & $e_p({\cal L}_p^{Bu})$& 
$e_p(10\, {\cal L}_p^{Bu})$ &  $e_p^{obs}$ \\
\hline
HD 209458b & $5.5\pm 1.5$\footnote{\citet{b2}} & 0.045 & 
$1.02\times10^{-8}$  & 0.014 & 0.046 &  $0.014\pm0.009$\\
WASP-1b  &$2.0\pm1.0$\footnote{\citet{s2}}  & 0.011 & $4.49\times10^{-8}$ & 0.012 & 0.037 &  -\\
HAT-P-1b & $3.6\pm 1.0^a$  & 0.109 &  $1.32\times10^{-8}$  & 0.031 & 0.098 & $0.09\pm0.02$\\
\hline
\end{tabular}
\end{minipage}
\end{table*}
\citet{b3} calculate ${\cal L}_p$ specifically for HD 209458b only (in addition
to various generic models), and find a value
for ${\cal L}_p$ ten times higher for a coreless planet, while
for a planet with a $20\, M_\oplus$ core, they find
${\cal L}_p$ is {\it another} factor of ten higher.
Note that they take a slightly
higher value for $R_p$ ($1.35\,R_J$).
Values for $e_p$ 
corresponding to ${\cal L}_p={\cal L}_p^{Bu}$
and $Q_p=10^5$ (see discussion below)
are shown in Table~\ref{theory} as well as values corresponding 
to ${\cal L}_p=10\,{\cal L}_p^{Bu}$. 
The estimated value of $e_p$ for HD 209458b (listed
under $e_p^{obs}$) is more
consistent with ${\cal L}_p={\cal L}_p^{Bu}$
than ${\cal L}_p=10\,{\cal L}_p^{Bu}$ 
while the opposite is true for HAT-P-1b,
although the eccentricity value for the latter awaits refinement.
Note that in reality ${\cal L}_p$ is a function of $R_p$ and $m_p$, however,
its role in equation~\rn{ept} is to parameterize various authors' models.

In both the models of \citet{b2} and \citet{b3}, tidal energy is deposited homogeneously 
throughout the planet's envelope and dissipated locally. If, for example,
tidal energy is predominantly 
dissipated in the {\it outer layers} 
of the planet, a significantly smaller value
of ${\cal L}_p$, and hence $e_p$, may be required to achieve
the same planetary radius (as long as the energy is dissipated
sufficiently deep below the surface; \cite{w2} finds that 
tidal excitation of inertial modes results in energy being dissipated
very close to the surface resulting in very little structural adjustment).
Alternatively, for a given $e_p$ one
would require larger values of $Q_p$ for correspondingly smaller
values of ${\cal L}_p$. Note that this scenario makes it difficult
to draw conclusions about the existence or otherwise of a planetary core.

There is significant uncertainty in the theory of tidal damping in stars and 
planets, much of it associated with internal structure and (in the case
of gaseous objects) the theory of turbulent dissipation
(see \citet{o1} for an excellent review).
In the case of solid planets,
one can infer damping timescales based on laboratory measurements
of $Q$-values of solid materials, and for the special case of the Earth-Moon
system, one can actually measure the rate of recession of the Moon.
In the case of Jupiter and its
satellite system, it is possible to put lower and upper bounds on Jupiter's
$Q$-value
($6\times 10^4-2\times 10^6$: Yoder \& Peale 1981).
These bounds are inferred from the existence of the Laplace resonance:
if $Q_J$ were too low Io would have already passed through the 2:1 resonance
with Europa (probably destabilizing the system),
while if it were too high it would never have been captured in the first place.
Similarly for Uranus, $Q_U<39,000$ \citep{t1}
while for Neptune, $12,000 < Q_N < 330,000$ \citep{b9}.

There are various factors which might be expected to influence the
$Q$-value of a hot Jupiter and how this might differ from that of Jupiter.
One obvious difference between hot Jupiters and our own Jupiter are
spin rates. While Jupiter's spin period is 10 hrs, hot Jupiters are approximately
synchronous with the orbital motion and hence have periods around
an order of magnitude longer. Goldreich \& Soter (1966, p388) point out
that the smaller the mass of a planet, the lower the $Q$-value. 
More recent work by \citet{o1} and \citet{w2} who consider 
tidal forcing in slowly rotating planets (appropriate to synchronously
rotating short-period planets) find that $Q$-values for such planets
tend to be similar to that of Jupiter. \citet{w3} shows that
the observed orbital parameters of short-period extrasolar planets
indicate that their $Q$-values lie between $3\times 10^4$ and
$3\times 10^5$. The lower value is increased if one
takes into account the fact that planets are likely to arrive at
their current
positions with significantly larger radii that those observed today
(see Fig.\ 1 of \citet{b4}). Throughout this paper we use mostly
$Q_p=10^5$ to illustrate the theory developed here.

\section{Long-term eccentricity evolution}
The estimated ages of HD 209459, HAT-P-1 and WASP-1
are shown in Table~\ref{theory}
together with the circularization timescales of their
planets (assuming $Q_p=10^5$).
Clearly a mechanism is needed to maintain orbital eccentricity
if tidal heating is the source of energy which maintains the inflated
planetary radii. 
It seems to be implicitly assumed that the oscillatory secular
variations in the eccentricities 
of a pair of close planets \citep{m3} prevail indefinitely,
even in the presence of tidal dissipation.
In this section we show that the natural endpoint of such evolution
is in fact a pair of circular orbits, with the structural properties of the tidally
forced planet determining (at least in part) the timescale on which
this occurs.
This evolution occurs in three stages (see Figure~\ref{eta-GR-tides}): (1) circulation of 
the angle between the lines of apsides of the two planets accompanied
by the slow oscillation of the eccentricities at constant amplitude,
together with decline
of the {\it mean} value of the inner planet's eccentricity until it reaches a (quasi)-fixed
value: this occurs on the circularization timescale; 
(2) libration of the angle between the lines of apsides
accompanied by the slow oscillation of the eccentricities with reducing 
amplitude but maintenance of the mean value of the inner eccentricity:
this occurs on {\it twice} the circularization timescale; and (3) either
(a): a slow non-oscillatory decline in both eccentricities to zero or
(b): maintenance of the inner eccentricity for some (often extremely long)
period of time followed by its {\it increase} to some maximum value
followed by an often fast decline of both eccentricities to zero. 
Whether or not the final phase occurs via route (a) or (b), it proceeds 
on a timescale sometimes several orders of magnitude longer than
the circularization timescale.

In general, the rates of change of a planet's orbital elements will have contributions
from companions if they exist, the relativistic post-Newtonian 
contribution to the potential of the star, the tidal and spin
bulges of the star and planet, and tidal dissipation in the atmospheres of the
star and the planet \citep{m1,a00,a0}. 
For coplanar systems, planetary elements which evolve are
the semimajor axis, the eccentricity and the longitude of periastron.
The only effect which contributes to a secular change in the semimajor axis
(except when the system is in a resonance)
is tidal dissipation, while for short-period systems
the apsidal motion is dominated by the third body and the relativistic 
potential of the star. The eccentricity in turn is dominated by the third
body and tidal damping. We will assume that the orbits are coplanar,
that the planet is synchronously
rotating with the orbital motion, and that the spin bulge of the star is
negligible. For the analysis we will assume that the rate
of change of the planet's semimajor axis is negligible,
while numerical integrations will include this effect (see \citet{r1}
and \citet{m4} for discussions of this point).

The plan of this section is as follows:
(1) Study the equations governing the secular evolution of three coplanar
Newtonian point masses for which $e_p$ is small compared to the
companion eccentricity and $m_p\ll m_*$,
and determine the regions of parameter space
for which the system librates or circulates. In particular, determine
the fixed points of this dynamical system;
(2) Determine how the results of (1) are modified by the inclusion
of post-Newtonian terms in the potential of the star;
(3) Introduce tidal damping and show that the system evolves towards
doubly circular orbits on timescales described above.

\subsection{Newtonian three-body secular evolution}

Consider a {\it coplanar} point-mass
three-body system with masses $m_*$, $m_p$ and $m_c$
(for star, planet and companion respectively) for which 
$m_p\ll m_*$, but with no constraints
on $m_c$, and let the star-planet and star-companion
semimajor axes be $a_p$ and $a_c$ respectively.\footnote{In fact,
a Jacobian coordinate system is used to derive equations~\rn{ep} to \rn{wc}
so that the companion's orbital elements are actually measured with respect
to the centre of mass of the star-planet system.}
Further, let the
corresponding eccentricities be $e_p$ and $e_c$, with $e_p\ll 1$ but again with
no constraint on $e_c$, and let the longitudes of periastra be
$\varpi_p$ and $\varpi_c$.
While no energy is exchanged between the orbits of a (nonresonant) 
stable system (and hence the semimajor axes remain constant), to first
order in $e_p$ 
the time-averaged rates of change of $e_p$, $e_c$, $\varpi_p$ and $\varpi_c$ are
(Mardling \& Lin (2002); 
also see \citet{m2} for the full secular equations to all orders and including 
inclination and precession)
\be
\frac{d e_p}{dt}=-\frac{15}{16}\,n_p\,e_c\,\left(\frac{m_c}{m_*}\right)
\left(\frac{a_p}{a_c}\right)^4\frac{\sin(\varpi_p-\varpi_c)}{(1-e_c^2)^{5/2}},
\label{ep}
\ee
\be
\frac{d e_c}{dt}=\frac{15}{16}\,n_c\,e_p\,\left(\frac{m_p}{m_*}\right)
\left(\frac{a_p}{a_c}\right)^3\frac{\sin(\varpi_p-\varpi_c)}{(1-e_c^2)^2},
\label{ec}
\ee
\be
\frac{d\varpi_p}{dt}=\frac{3}{4}n_p\left(\frac{m_c}{m_*}\right)
\left(\frac{a_p}{a_c}\right)^3(1-e_c)^{-3/2}
\left[1-\frac{5}{4}\left(\frac{a_p}{a_c}\right)\left(\frac{e_c}{e_p}\right)
\frac{\cos(\varpi_p-\varpi_c)}{1-e_c^2}\right]
\label{wp}
\ee
and
\be
\frac{d\varpi_c}{dt}=\frac{3}{4}n_c\left(\frac{m_p}{m_*}\right)
\left(\frac{a_p}{a_c}\right)^2
\left(1-e_c^2\right)^{-2}\left[
1-\frac{5}{4}\left(\frac{a_p}{a_c}\right)
\left(\frac{e_p}{e_c}\right)\frac{\left(1+4e_c^2\right)}{\left(1-e_c^2\right)}
\cos(\varpi_p-\varpi_c)\right].
\label{wc}
\ee
where $n_p$ and $n_c$ are the orbital frequencies (mean motions) of the
planet and companion respectively.
 Note that
\citet{m1} and \citet{m2} employ a Legendre expansion in the 
ratio $a_p/a_c$ to derive the 
rates of change of the elements and make no assumptions about
the mass ratios, eccentricities or inclinations, in contrast to the {\it literal} expansion
used by \citet{m3} which relies on the smallness of 
the eccentricities and inclinations but holds for all semimajor axes ratios.
Equations~\rn{ep} to \rn{wc} are given to {\it octopole} order (corresponding
to the fourth power in $a_p/a_c$);
see also \citet{l2} and \citet{f2}.

\subsubsection{Libration versus Circulation of $\varpi_p-\varpi_c$}\label{libcirc}

\begin{figure}
\includegraphics[width=150mm]{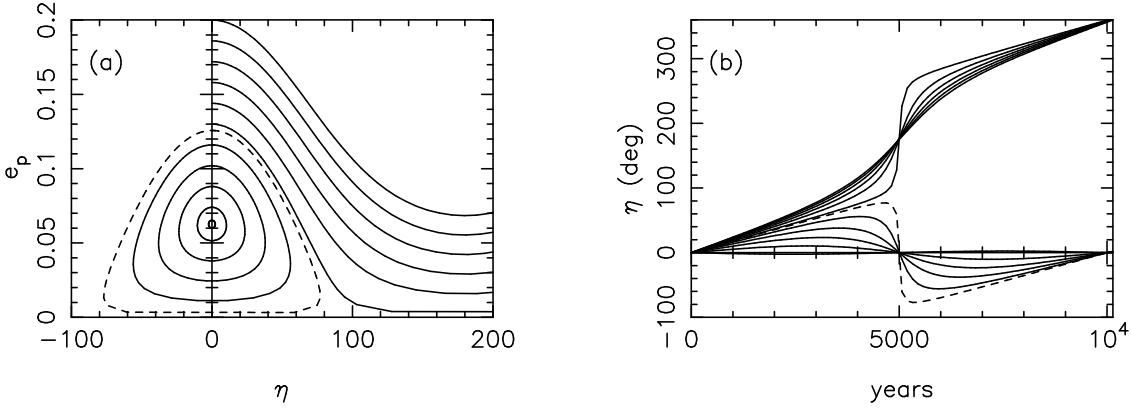}
\caption{Evolution of $e_p$ and $\eta$ for an HD 209458 type system (parameters
given in text). The full secular equations were integrated to produce these curves.
(a) shows libratory and circulatory behaviour consistent with
$e_p^{(eq)}=0.063$ and $e_p^{boundary}=0.126$, while
(b) demonstrates that $\dot\eta$ is almost constant for $e_p(0)$ not
too close to $e_p^{boundary}$. The dashed curve in both plots
corresponds to $e_p(0)=e_p^{boundary}$. The libration and circulation
periods are consistent with the analytical value of $1.01\times 10^4$ yr.}
\label{etafig}
\end{figure}

In order to derive simple conditions under which the angle
$\varpi_p-\varpi_c$ circulates or librates, we assume that
$e_c$ is significantly greater than $e_p\ll1$ (a situation
relevant to the present study; see \citet{n2} for an
alternative treatment). Then from \rn{ec}, $\dot{e}_c\simeq 0$, and
we can write
\be
\dot{e}_p=-W_o\,e_c\,\sin\eta
\label{ep2}
\ee
and
\be
\dot\eta=W_q-W_o\left(\frac{e_c}{e_p}\right)\cos\eta,
\label{eta}
\ee
where $\eta\equiv \varpi_p-\varpi_c$ and
$W_q$ and $W_o$ are constants ($q$ for quadrupole and $o$ for
octopole) defined by
\be
W_q=\frac{3}{4}n_p \left(\frac{a_p}{a_c}\right)^3
\left(\frac{m_c}{m_*}\right)\varepsilon_c^{-3}
\left[1-\sqrt{\frac{a_p}{a_c}}\left(\frac{m_p}{m_c}\right)\varepsilon_c^{-1}\right]
\label{Wq}
\ee
and
\be
W_o=\frac{15}{16}n_p\left(\frac{a_p}{a_c}\right)^4
\left(\frac{m_c}{m_*}\right)\varepsilon_c^{-5},
\ee
where $\varepsilon_c=\sqrt{1-e_c^2}$.
The fixed points of \rn{ep2} and \rn{eta} are 
\be
(e_p,\eta)=(e_cW_o/W_q,2k\pi)\,\,\,
{\rm when}\,\,\, W_q>0,
\ee
and 
\be
(e_p,\eta)=(-e_cW_o/W_q,(2k+1)\pi)\,\,\, 
{\rm when}\,\,\ W_q<0.
\ee
Here $k$ is any integer.
Writing $x\equiv e_p-e_p^{(eq)}$, where 
\be
e_p^{(eq)}=e_cW_o/|W_q|=\frac{(5/4)(a_p/a_c)\,e_c\,\varepsilon_c^{-2}}
{\left|1-\sqrt{a_p/a_c}(m_p/m_c)\varepsilon_c^{-1}\right|},
\label{equil}
\ee
and linearizing about $\eta=0$
we have
\be
\dot x=-e_cW_o\,\eta, \hspace{0.5cm}
\dot\eta=\left(\frac{W_q^2}{W_o}\right)\,\frac{x}{e_c},
\label{x}
\ee
or
\be
\ddot\eta+W_q^2\,\eta=0
\label{etadd}
\ee
which has solutions
\be
\eta(t)=A\,\cos(W_q t+\beta),
\label{eta2}
\ee
where
\be
A^2=\eta(0)^2+\left(1-e_p(0)/e_p^{(eq)}\right)^2
\hand
\tan\beta=\left(1-e_p(0)/e_p^{(eq)}\right)/\eta(0).
\label{Abeta}
\ee
From \rn{etadd} we conclude that the fixed point $\eta=0$ is stable.
For libration around $2k\pi$, $k\neq 0$, add $2k\pi$ to \rn{eta2}.
From \rn{x} we have that
\be
e_p(t)=e_p^{(eq)}\left[1-A\,\sin(W_q t+\beta)\right]\equiv e_p^{(lib)}(t)
\label{ep3}
\ee
so that the maximum and minimum values of $e_p$ when a system is librating 
are approximately
\be
e_p^{max,min}=\left(1\pm A\right)e_p^{(eq)}.
\label{epmax}
\ee
For libration about $\eta=\pi$, also a stable fixed point,
\be
\eta(t)=\pi-A\,\cos(W_q t+\beta),
\label{eta3}
\ee
with $A$ and $\beta$, $e_p(t)$ and  $e_p^{max,min}$ again given
by \rn{Abeta},
\rn{ep3} and \rn{epmax} respectively.
For libration around $(2k+1)\pi$, $k\neq 0$, add $2k\pi$ to \rn{eta3}.
Since $W_q>0$ for libration about $\eta=0$ and $W_q<0$ for libration 
about $\eta=\pi$, from \rn{Wq}
we have that
\be
\frac{m_c}{m_p}>\sqrt{\frac{a_p}{a_c}}\varepsilon_c^{-1}\,\,
{\rm for\,\, libration\,\, about}\,\, \eta=0;\hspace{0.5cm}
\frac{m_c}{m_p}<\sqrt{\frac{a_p}{a_c}}\varepsilon_c^{-1}\,\,
{\rm for\,\, libration\,\, about}\,\, \eta=\pi.
\ee
Consider libration about $\eta=0$.
This can occur as long as there are solutions to $\dot\eta=0$,
that is, there exist real values for $\eta$ when $\cos\eta=(e_p/e_c)\,W_q/W_o$.
From \rn{ep3} and \rn{equil} this is (in the linear approximation)
\be
\cos\eta=1-A\,\sin(W_q t+\beta)
\ee
whose minimum value is $1-A$ (taking $A$ to be positive)
corresponding to the maximum value $\eta$
achieves during a libration cycle. But in order for $e_p$ to be positive or zero,
$0\leq A\leq 1$ from \rn{epmax}. Thus we have the result that libration 
about $\eta=0$ can
occur as long as $|\eta|<\pi/2$. Moreover, from \rn{epmax} 
the maximum value of $e_p$ associated with $A=1$, and hence the 
value associated with the boundary between libration and circulation, is
\be
e_p^{boundary}=2\,e_p^{(eq)}
\ee
which also holds for libration about $\eta=\pi$. In the latter case libration can
occur as long as $|\eta-\pi|<\pi/2$. 
These results are consistent with 
numerically integrating the full equations (equations \rn{ep} to \rn{wc} but
with the full dependence on $e_p$; see \citet{m2}), an example of which
is presented in
Figure~\ref{etafig}
for a system with
$m_p=0.64\, M_J$,
$m_c=1\,M_J$, $m_*=1.14\,M_\odot$, $a_p=0.047$ AU, $a_c=0.4$ AU
and $e_c=0.3$. The values for the inner planet are appropriate for
HD 209458b. Figure~\ref{etafig}(a)
plots $e_p$ against $\eta$ for $\eta(0)=0$ and 12 values of $e_p(0)$
ranging from 0.06 to 0.2. Equation~\rn{equil} gives $e_p^{(eq)}=0.063$ and
$e_p^{boundary}=0.126$.
Figure~\ref{etafig}(b) shows that
for $e_p>e_p^{boundary}$, $\eta$ circulates with a frequency that is
approximately constant ($\simeq W_q$) as long as $e_p$ is not
too close to $e_p^{boundary}$.
Putting $\dot\eta=W_q$ gives
\be
\eta(t)=W_q\,t+\eta(0),
\label{etacirc}
\ee
so that from \rn{ep2}, $\dot e_p=-W_o\sin(W_q t+\eta(0))$ and
\be
e_p(t)=e_p(0)+(W_o/W_q)\left[\cos(W_q t+\eta(0))-\cos\eta(0)\right]
\equiv e_p^{(circ)}(t).
\label{ep4}
\ee
Therefore if $W_q>0$, the maximum and minimum values of 
$e_p$ correspond to $\eta=2k\pi$ and $\eta=(2k+1)\pi$ respectively, where
$k$ is an integer, and are
given by
\be
e_p^{max,min}=e_p(0)\pm e_p^{(eq)}\left[1\mp\cos\eta(0)\right],
\label{maxmin0}
\ee
while for $W_q<0$, these are
\be
e_p^{max,min}=e_p(0)\pm e_p^{(eq)}\left[1\pm\cos\eta(0)\right]
\label{maxminpi}
\ee
corresponding to $\eta=(2k+1)\pi$ and $\eta=2k\pi$ respectively.
From \rn{ep3} and \rn{ep4} we see that both the libration and
circulation periods are given by $2\pi/W_q$.
For the example shown in Figure~\ref{etafig}, $2\pi/W_q=1.01\times 10^4$ yr.

The actual time variation of $e_c$ can be obtained by forming $de_p/de_c$
from \rn{ep} and \rn{ec}, which upon integration gives
\be
e_c(t)=\left[1-\left(\beta\,e_p(t)^2-C\right)^2\right]^{1/2},
\label{ect}
\ee
where 
\be
\beta=\frac{1}{2}\sqrt{\frac{a_p}{a_c}}\left(\frac{m_p}{m_c}\right)
\hand
C=\beta\,e_p(0)^2-\sqrt{1-e_c(0)^2},
\ee
and $e_p(t)$ is given by \rn{ep3} or \rn{ep4} according to whether
$\eta$ librates or circulates.

Summarizing these results, we have that libration will occur about the
fixed point $(e_p,\eta)=(e_p^{(eq)},0)$ at frequency $W_q>0$ when
\be
\frac{m_c}{m_p}>\sqrt{\frac{a_p}{a_c}}\varepsilon_c^{-1}
\hand
0\leq e_p<2\,e_p^{(eq)}
\hand
-\pi/2<\eta<\pi/2,
\label{lib0}
\ee
while libration will occur about the
fixed point $(e_p,\eta)=(e_p^{(eq)},\pi)$ at frequency $W_q<0$ when
\be
\frac{m_c}{m_p}<\sqrt{\frac{a_p}{a_c}}\varepsilon_c^{-1}
\hand
0\leq e_p<2\,e_p^{(eq)}
\hand
\pi/2<\eta<3\pi/2.
\label{libpi}
\ee
Similar results hold for $k\neq 0$.
If neither of the conditions \rn{lib0} or \rn{libpi} are met, $\eta$ will 
circulate at frequency $W_q$ 
with the maximum and minimum values of $e_p$ given
by \rn{maxmin0} when $m_c/m_p>\varepsilon_c^{-1}\sqrt{a_p/a_c}$ ($W_q>0$),
and  \rn{maxminpi} when $m_c/m_p<\varepsilon_c^{-1}\sqrt{a_p/a_c}$ ($W_q<0$).

\subsection{Relativistic Effects}

\begin{figure}
\includegraphics[width=150mm]{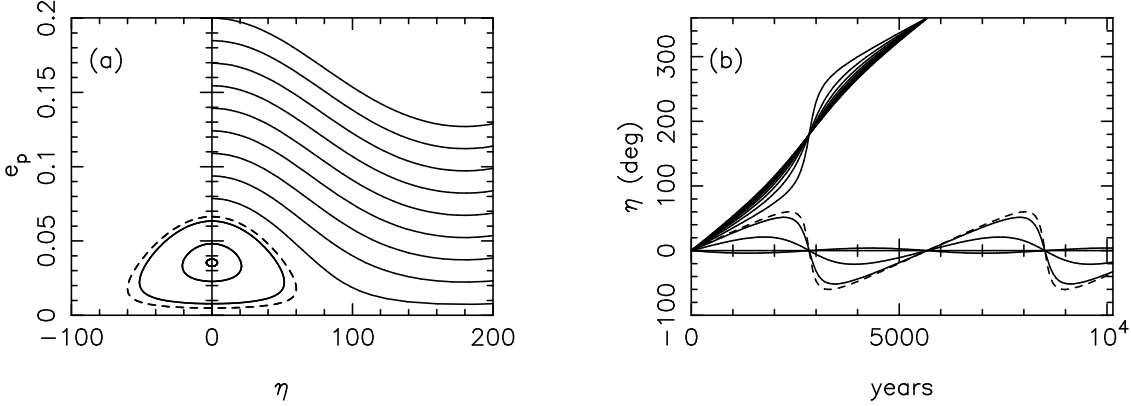}
\caption{Same as Figure~\rn{etafig} but with $W_{GR}$ included.
Libration and circulation of $\eta$ is more rapid and the maximum amplitude
for librating systems is reduced in this case. The dashed curves correspond
to $e_p=e_p^{boundary}$. The libration and circulation
periods are consistent with the analytical value of 5700 yr.
}
\label{eta-GR}
\end{figure}

Contributions to the potential of the star from post-Newtonian terms
affect the apsidal advance only. The orbit-averaged contribution to $\dot\eta$ from
the innermost orbit is
(eg.\ Mardling \& Lin, 2002)
\be
W_{GR}=\frac{3 n_p}{1-e_p^2}\left(\frac{n_p\,a_p}{c}\right)^2,
\ee
where $c$ is the speed of light,
so that
\be
\dot\eta=W_q+W_{GR}-W_o\left(\frac{e_c}{e_p}\right)\,\cos\eta
\equiv
\oW-W_o\left(\frac{e_c}{e_p}\right)\,\cos\eta,
\label{etaGR}
\ee
where 
\be
\oW=\frac{3}{4}n_p \left(\frac{a_p}{a_c}\right)^3
\left(\frac{m_c}{m_*}\right)\varepsilon_c^{-3}
\left[1-\sqrt{\frac{a_p}{a_c}}\left(\frac{m_p}{m_c}\right)\varepsilon_c^{-1}
+\gamma\varepsilon_c^3\right],
\label{oW}
\ee
and $\gamma=4(n_p a_p/c)^2(m_*/m_c)(a_c/a_p)^3$ 
is, to first-order in 
$e_p$, the ratio of $W_{GR}$ to
the quadrupole contribution to $\dot\varpi_p$.
One effect of $W_{GR}$ on the system is to shorten the 
eccentricity modulation and apsidal period
(compare Figure~\ref{eta-GR} with Figure~\ref{etafig}). 
In this example, $W_{GR}\simeq W_q$ in contrast to Mercury's
orbit around the Sun for which $W_{GR}$ is only 7\% of the 
contributions to $\oW$ from the other planets in the Solar System.
The equilibrium eccentricity is now given by\footnote{Note that 
equation~(4) in Mardling \& Lin (2004) is in error by a factor of 2.}
\be
e_p^{(eq)}=e_cW_o/|\oW|=\frac{(5/4)(a_p/a_c)\,e_c\,\varepsilon_c^{-2}}
{\left|1-\sqrt{a_p/a_c}(m_p/m_c)\varepsilon_c^{-1}+\gamma\varepsilon_c^3\right|}.
\label{equilGR}
\ee
Thus when $\sqrt{a_p/a_c}(m_p/m_c)\varepsilon_c^{-1}<1$,
the effect of $W_{GR}$ is to decrease the equilibrium eccentricity,
an example of which is shown in Figure~\ref{eta-GR},
while for $\sqrt{a_p/a_c}(m_p/m_c)\varepsilon_c^{-1}> 1$
the opposite is true.
\citet{m4} show that the former can be vital for the survival
of short period terrestrial planets with companions.

The analogues of \rn{lib0} and \rn{libpi} are that
libration will occur about the
fixed point $(e_p,\eta)=(e_p^{(eq)},0)$ at frequency $\oW>0$, where
$e_p^{(eq)}$ is now given by \rn{equilGR}, when
\be
\frac{m_c}{m_p}>\sqrt{\frac{a_p}{a_c}}\varepsilon_c^{-1}/(1+\gamma\varepsilon_c^3)
\hand
0\leq e_p<2\,e_p^{(eq)}
\hand
-\pi/2<\eta<\pi/2,
\label{lib0r}
\ee
while libration will occur about the
fixed point $(e_p,\eta)=(e_p^{(eq)},\pi)$  at frequency $\oW<0$ when
\be
\frac{m_c}{m_p}<\sqrt{\frac{a_p}{a_c}}\varepsilon_c^{-1}/(1+\gamma\varepsilon_c^3)
\hand
0\leq e_p<2\,e_p^{(eq)}
\hand
\pi/2<\eta<3\pi/2.
\label{libpir}
\ee
If neither of the conditions \rn{lib0r} or \rn{libpir} are met, 
then when 
$m_c/m_p>\varepsilon_c^{-1}\sqrt{a_p/a_c}/(1+\gamma\varepsilon_c^3)$,
$\eta$ will 
circulate at frequency $\oW$ in the positive direction 
with the maximum and minimum values of $e_p$ given
by \rn{maxmin0}, while when
$m_c/m_p<\varepsilon_c^{-1}\sqrt{a_p/a_c}/(1+\gamma\varepsilon_c^3)$,
$\eta$ will circulate in the negative direction at the same frequency
with maximum and minimum values of $e_p$ given by \rn{maxminpi}.
The time variations of $\eta$, $e_p$ and $e_c$ are given by
the equations set out in the previous section, with $e_p^{(eq)}$
given by \rn{equilGR}.

The form of \rn{equilGR} suggests that for a range of parameters,
$e_p^{(eq)}$ can be large. Singular values of $e_p^{(eq)}$ exist for some combinations
of parameters, although when the octopole term in \rn{wc} is included in \rn{eta}
these are replaced by finite, although still relatively large values (see \citet{n2}).
As we will show in Section~\ref{incep}, this has important consequences
for eccentricity forcing by low-mass
companions.

\subsection{Tidal damping and quasi-fixed-point behaviour}\label{dissipation} 
While tidal dissipation in the innermost planet directly affects
the rate of change of its eccentricity, it also indirectly affects
the rate of change of the companion planet's eccentricity
(recall that the previous analysis assumed $e_c$ was constant),
as well as the angle between the apsidal lines, $\eta$.\footnote{The 
contributions to $\dot\eta$
from the tidal and spin bulges of the star and planets
are negligible compared to the terms included here.}
The relevant system of equations now becomes
\be
\dot{e}_p=-W_o\,e_c\,\sin\eta-W_T\,e_p,
\label{ep3T}
\ee
\be
\dot{e}_c=W_c\,e_p\,\sin\eta,
\label{ec3T}
\ee
\be
\dot\eta=\overline{W}_q-W_o\left(\frac{e_c}{e_p}\right)\cos\eta,
\label{eta3T}
\ee
where from \rn{tcirc},
\be
W_T=\frac{21}{2}\,n_p\left(\frac{k_p}{Q_p}\right)
\left(\frac{m_*}{m_p}\right)\left(\frac{R_p}{a_p}\right)^5=\tau_{circ}^{-1}
\label{epT}
\ee
and from \rn{wc},
\be
W_c=\frac{15}{16}\,n_c\,\left(\frac{m_p}{m_*}\right)
\left(\frac{a_p}{a_c}\right)^3\varepsilon_c^{-4},
\ee
where again we have ignored the octopole term (proportional to 
$e_p/e_c$) in \rn{wc}.
The damped autonomous
system \rn{ep3T}-\rn{eta3T}
evolves to the formal fixed point $(e_p,e_c,\eta)=(0,0,\eta^*)$,
where $\eta^*$ is given below and is such that $|\eta^*|({\rm mod}\,2\pi)\ll0$
or $|\eta^*-\pi|({\rm mod}\,2\pi)\ll0$ for $\oW>0$ and $\oW<0$ respectively.
However, there are three timescales associated with this evolution:
(1) The system circulates until $e_p$, $e_c$ and $\eta$ are such that
libration conditions are met (see \rn{lib0} and \rn{libpi}:
the system is ``captured'' by the associated fixed point).
This occurs on the circularization
timescale of the inner orbit,\footnote{This
phase is skipped if the initial conditions are 
such that the system starts librating immediately.}
 $\tau_{circ}$;
(2) The system librates about and evolves towards the
(now quasi-)fixed point $(e_p,e_c,\eta)=(e_p^*,e_c^*,\eta^*)$, 
where $e_c^*\simeq e_c(0)$
and $e_p^*\simeq e_p^{(eq)}$ (equation \rn{equilGR}),
on the timescale $2\,\tau_{circ}$ ($e_c^*$ and $e_p^*$ will be properly
defined below);
(3) The system evolves smoothly
towards the formal fixed point
$(e_p,e_c,\eta)=(0,0,\eta^*)$ on some timescale $\tau_c\gg\tau_{circ}$.
The details for these phases are as follows, where for the first two we assume that
$\dot e_c\ll \dot e_p$ so that $e_c\simeq e_c(0)$.

\subsubsection{Circulation}
Following Section~\ref{libcirc}, the circulation phase can be approximated
with $\dot\eta=\oW$. Then \rn{ep3T} becomes
\be
\dot e_p+W_T\,e_p=-W_o\,e_c\,\sin(\oW t+\eta(0))
\ee
which has solution
\be
e_p(t)
=C\,e^{-W_T t}+\sigma e_p^{(eq)}\cos(\oW t+\eta(0))
\label{edecline}
\ee
where $\sigma={\rm sgn}\left[\oW\right]$,
$C=e_p(0)-\sigma e_p^{(eq)}\cos\eta(0)$
and we have assumed that $W_T\ll |\oW|$.
This phase is shown in blue in Figure~\ref{eta-GR-tides}
and evolves on the timescale $1/W_T=\tau_{circ}$.
Notice that the amplitude of variation is not affected by
dissipation and is given by $ e_p^{(eq)}$. Also note that $Q_p$
is set artificially low in order to clearly demonstrate how this phase
proceeds. 

\subsubsection{Libration}
The quasi-fixed point $(e_p^*,e_c^*,\eta^*)$ is obtained by
putting $\dot e_p=0$ and $\dot\eta=0$ in \rn{ep3T} and \rn{eta3T}
respectively. While there is only one true equilibrium value of $e_c$
(that is, $e_c=0$), we take $e_c^*$ to be the value of $e_c$ when
the libration amplitude of $e_p$ drops below some specified small value
(see next section).
Eliminating $e_p/e_c$ between \rn{ep3T} and \rn{eta3T} then gives
\be
\sin\eta^*
=-\frac{W_T}{\oW}\cos\eta^*\simeq
-\frac{W_T}{|\oW|}<0,
\label{etastar}
\ee
where the second step comes from the fact that
${\rm sgn}(\cos\eta^*)={\rm sgn}(\oW)$ and $W_T/|\oW|\ll1$.
Thus for $\oW>0$, $\eta^*({\rm mod}\,2\pi)\simeq -W_T/\oW$ while
for $\oW<0$, $(\eta^*-\pi)({\rm mod}\,2\pi)\simeq -W_T/\oW$.
For HD 209458b with a $0.1 M_J$ companion at 0.4 AU,
$\eta^*= 10^{-5}-10^{-4}$ radians, depending on the value of $e_c$ 
(taking $Q_p=10^5$).
The value of $e_p^*$ is given by \rn{equilGR} with 
$e_c=e_c^*$.
Linearizing
about $(e_p^*,e_c^*,\eta^*)$
and putting $x=e_p-e_p^*$ and $y=\eta-\eta^*$ produces
the system
\be
\dot x=-e_c^* W_o\,y-W_T\, x,
\ee
\be
\dot y=(\oW^2/e_c^*W_o)\,x,
\ee
which yields the damped librating solution
\be
e_p(t)=e_p^*+A\,e^{-W_T t/2}\cos(W_q t+\delta)
\label{eplib}
\ee
and
\be
\eta(t)=\eta^*-A\,e^{-W_T t/2}\sin(W_q t+\delta).
\label{etalib}
\ee
For the purposes of this analysis, the constants $A$ and $\delta$ 
can be taken to be such that the librating phase is entered at the
first occurrence of $e_p=0$.
Writing the time at which this occurs as $t=t_{lib}$, 
\rn{eplib} and \rn{etalib} become
\be
e_p(t)=e_p^*\left\{1-e^{-W_T (t-t_{lib})/2}\cos\left[W_q (t-t_{lib})\right]\right\}
\ee
and
\be
\eta(t)=\eta^*-e^{-W_T (t-t_{lib})/2}\sin\left[W_q (t-t_{lib})\right]
\ee
In contrast to the circulating phase, 
the amplitude of oscillation of 
$e_p(t)$ diminishes on a timescale $2/W_T=2\,\tau_{circ}$ as the system
approaches the quasi-fixed point.
Figure~\ref{eta-GR-tides} shows this behaviour in red.
\begin{figure}
\includegraphics[width=140mm]{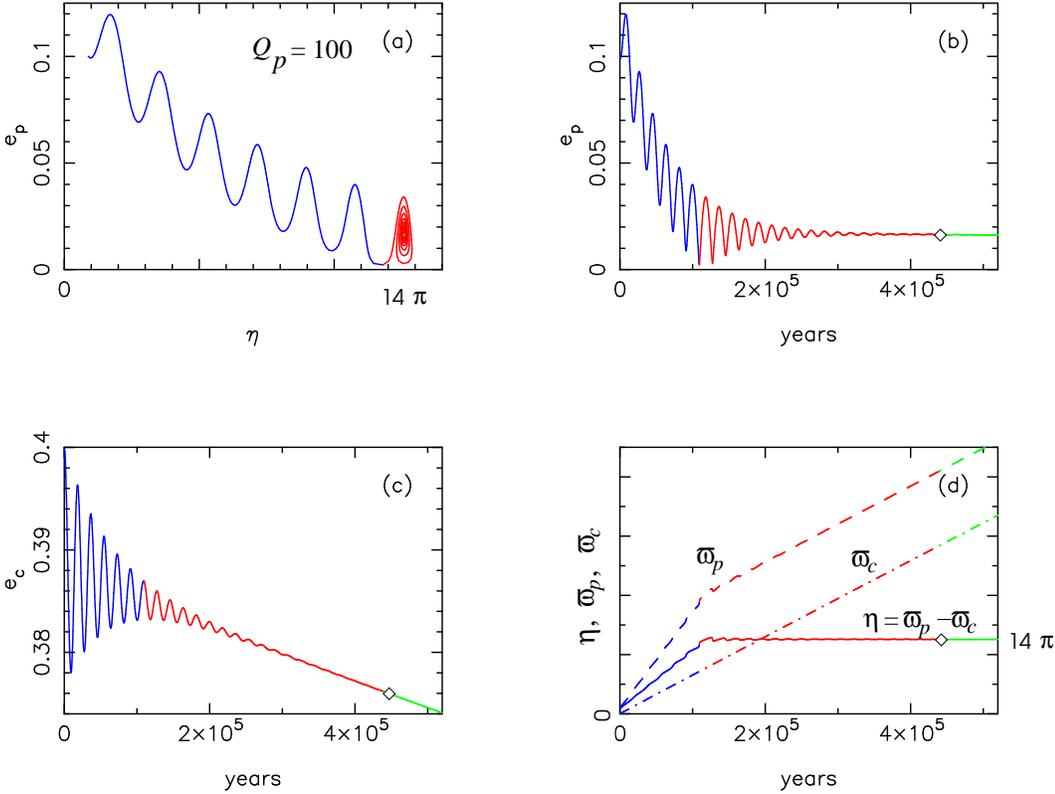}
\caption{Capture phase for HD 209458b with a $0.1 M_J$ companion
at 0.4 AU with $e_c=0.4$. The full secular equations including
evolution of $a_p$ and $a_c$ are integrated \citep{m4}.
The $Q$-value of the observed planet
is set artificially low at $Q_p=100$ in order to illustrate the process
(timescales are linearly proportional to $Q_p$ and values for
$e_c^*$ are not very sensitive to the value of $Q_p$; for this example,
the relative change in $e_c^*$ is less than 2\% over 
four orders of magnitude in $Q_p$). Blue curves correspond
to the circulation phase, red to the libration phase and green to the 
slow evolution or quasi-fixed point 
phase (see text for more discussion). In panel (c) notice that
the relative change in $e_c$ is small, justifying
(at least for this example) taking $e_c\simeq$ const during this phase
in the analysis.
The diamond symbols correspond to the points $(t_c,e_p^*)$, $(t_c,e_c^*)$
and $(t_c,\eta^*)$ in panels (b), (c) and (d) respectively.
Panel (d) shows the apsidal advance of the inner and outer orbits, with
$\dot\varpi_p$ locking onto $\dot\varpi_c$. This occurs irrespective
of the mass of the outer body because $e_p/e_c\ll 1$ (see equation~\rn{wc}).
}
\label{eta-GR-tides}
\end{figure} 

For the analysis above we have assumed $e_c$ is constant. However,
as Figure~\ref{eta-GR-tides}(c) illustrates, $e_c$ slowly declines
and has no analogue of $e_p^{(eq)}$ (except $e_c=0$).
The following section studies the effect of this on the long-term evolution
of the system.

\subsubsection{Long-term evolution}\label{longterm}

Once the quasi-fixed point is reached, the companion's eccentricity
evolves according to \rn{ec3T} with $e_p=e_cW_o/|\oW|$ 
and $\sin\eta=-W_T/|\oW|$:
\bea
\dot e_c&=&-\frac{W_cW_oW_T}{\oW^2}\,e_c\next
&=&
-\left(\frac{25}{16}\right)\left(\frac{m_p}{m_c}\right)\left(\frac{a_p}{a_c}\right)^{5/2}W_T
\cdot \frac{e_c}{F(e_c)}\next
&=& -\left(\frac{\lambda}{\tau_{circ}}\right)\frac{e_c}{F(e_c)}
\label{eclong}
\eea
where
\be
\lambda=
\left(\frac{25}{16}\right)\left(\frac{m_p}{m_c}\right)\left(\frac{a_p}{a_c}\right)^{5/2}
\hand
F(e_c)=\varepsilon_c^{3}(1-\alpha\varepsilon_c^{-1}
+\gamma\,\varepsilon_c^3)^{2}
\ee
with $\alpha=\sqrt{a_p/a_c}(m_p/m_c)$ (recall that $\varepsilon_c=\sqrt{1-e_c^2}$).
Note that $\alpha$ and $\gamma$ are effectively constant throughout
the evolution since the semimajor axes evolve on timescales much
longer than the age of the system (see Figure~\ref{eup2}(d)).
One can obtain an approximate timescale on which $e_c$ (and hence
$e_p$) evolves by putting 
$\varepsilon_c\simeq\varepsilon(e_c^*)\equiv\varepsilon_c^*$, where $e_c^*$ is yet to be formally defined.
Equation \rn{eclong} then gives
\be
e_c(t)=e_c(t_c)\,e^{-(t-t_c)/\tau_c},
\label{ecapprox}
\ee
where 
\be
\tau_c=
\left(F(e_c^*)/\lambda\right)\tau_{circ} 
\label{tauc}
\ee
and $t_c$ corresponds to the beginning of the slow evolution phase
which we choose to be
the first instance of $|\eta-\eta^*|<\epsilon$ for all $t>t_c$.
Here $\epsilon\ll1$ is an arbitrarily chosen parameter and is
taken to be $\epsilon=10^{-4}$ for the studies presented in this paper.
This then
serves as definitions of $e_c^*$ and $e_p^*$, that is, $e_c^*=e_c(t_c)$
and $e_p^*=e_p(t_c)$. 
The transition to the slow phase at 
time $t_c$ is indicated in Figures~\ref{eta-GR-tides}(b)-(d) by a diamond
symbol.
Note that $\tau_c\gg\tau_{circ}$ for most stable systems, except when
$F(e_c)$ is small (see Section~\ref{incep}).
The approximate solution \rn{ecapprox} is the
blue dot-dashed curve in Figure~\ref{eta-GR-tides-long}(a).
\begin{figure}
\includegraphics[width=150mm]{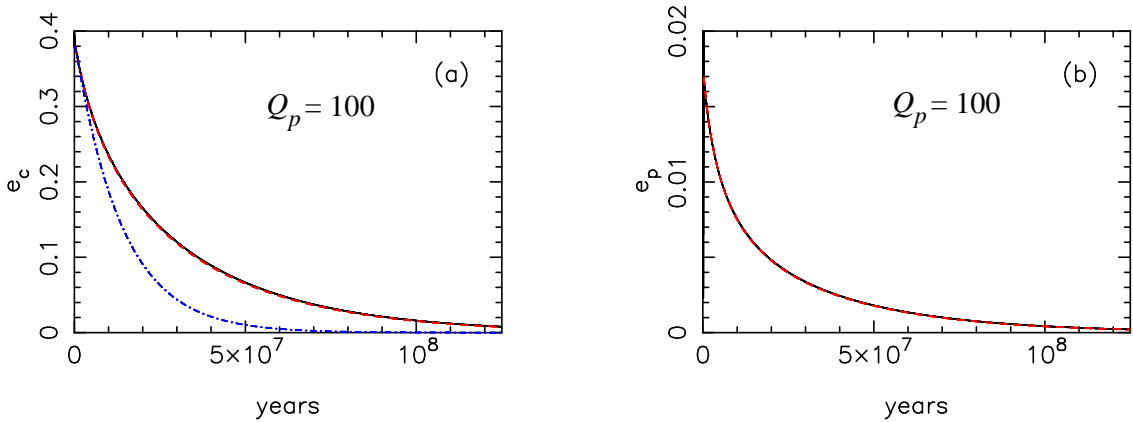}
\caption{Continuation of numerical solution in Figure~\ref{eta-GR-tides} 
showing the long-term evolution of (a) $e_c$ and (b) $e_p$ (black
curves, sitting underneath red curves), again with $Q_p=100$ for illustration
(note the $y$-axis scales are different to those in Figure~\ref{eta-GR-tides}). 
This corresponds to 125 Gyr of evolution
when $Q_p=10^5$ (compare this to $\tau_{circ}=0.045$~Gyr).
Also plotted in (a) is approximate solution \rn{ecapprox} (blue dot-dash curve)
and the analytic solution \rn{ecexact} (dashed red curve).
The approximate solution underestimates the timescale on which 
$e_c$ evolves by a factor of 1.8.
Also plotted in (b) is the analytic expression for $e_p(t)$ (red curve)
using \rn{equilGR} with $e_c=e_c(t)$ from \rn{ecexact}.
The analytic solutions are 
essentially indistinguishable from the numerical solutions.
}
\label{eta-GR-tides-long}
\end{figure} 

Keeping the full dependence on $e_c$ in \rn{eclong} and integrating
gives
\be
f(e_c;\alpha,\gamma)-f(e_c^*;\alpha,\gamma)=-\lambda(t-t_c)/\tau_{circ},
\label{ecaccurate}
\ee
where
\bea
f(x;\alpha,\gamma)&=&-\ff{1}{3}\gamma x^6+\ff{3}{2} \gamma
   x^4+(\alpha-3 \gamma)\, x^2+
   \sqrt{1-x^2} \left[\alpha^2-\ff{42}{315} \gamma
   \left(3 x^4-11 x^2+23\right)
   \alpha-\ff{1}{3} \left(x^2-4\right)+\gamma^2
  g(x)\right]
  \next
&&  
+(\gamma-\alpha+1)^2\ln x-\left[(\gamma-\alpha)^2+1\right]\ln\left(1+\sqrt{1-x^2}\right)
\label{ecexact}
\eea
with
\be
g(x)= \left(35 x^8-185 x^6+408 x^4-506
   x^2+563\right)/315.
\ee
This solution is shown as a dashed red curve in 
Figure~\ref{eta-GR-tides-long}(a) and is clearly
superior to the estimate \rn{ecapprox} which {\it underestimates}
the timescale on which the system evolves once apsidal locking
has taken place. We define the true timescale on which 
circularization of the companion orbit takes place (the $e$-folding time) to be the time
corresponding to $e_c^*/e$, that is, from \rn{ecaccurate},
\be
\tau_c^{true}=\left[3+\lambda^{-1}
\left(f(e_c^*;\alpha,\gamma)-f(e_c^*/e;\alpha,\gamma)\right)\right]\tau_{circ},
\label{tautrue}
\ee
where we have put $t_c=3\,\tau_{circ}$ (1 for the circulation phase
+ 2 for the libration phase).
As $e_c$ slowly decreases, $e_p$ maintains a value given by
\rn{equilGR}. Any departure from this quasi-equilibrium value
is quickly damped out on the circularization timescale $\tau_{circ}$.
Thus we can use \rn{ecaccurate} and  \rn{equilGR}
to calculate analytically $e_c(t)$ and $e_p(t)$ during the slow phase,
and hence to determine the current value of $e_p$ using the estimated age
of the system.
The analytic solution for $e_p(t)$ 
is shown in Figure~\ref{eta-GR-tides-long}(b) (red curve) and is 
indistinguishable from the numerical solution.

\subsubsection{The increasing $e_p$ regime: very low-mass companions}\label{incep}

\begin{figure}
\includegraphics[width=150mm]{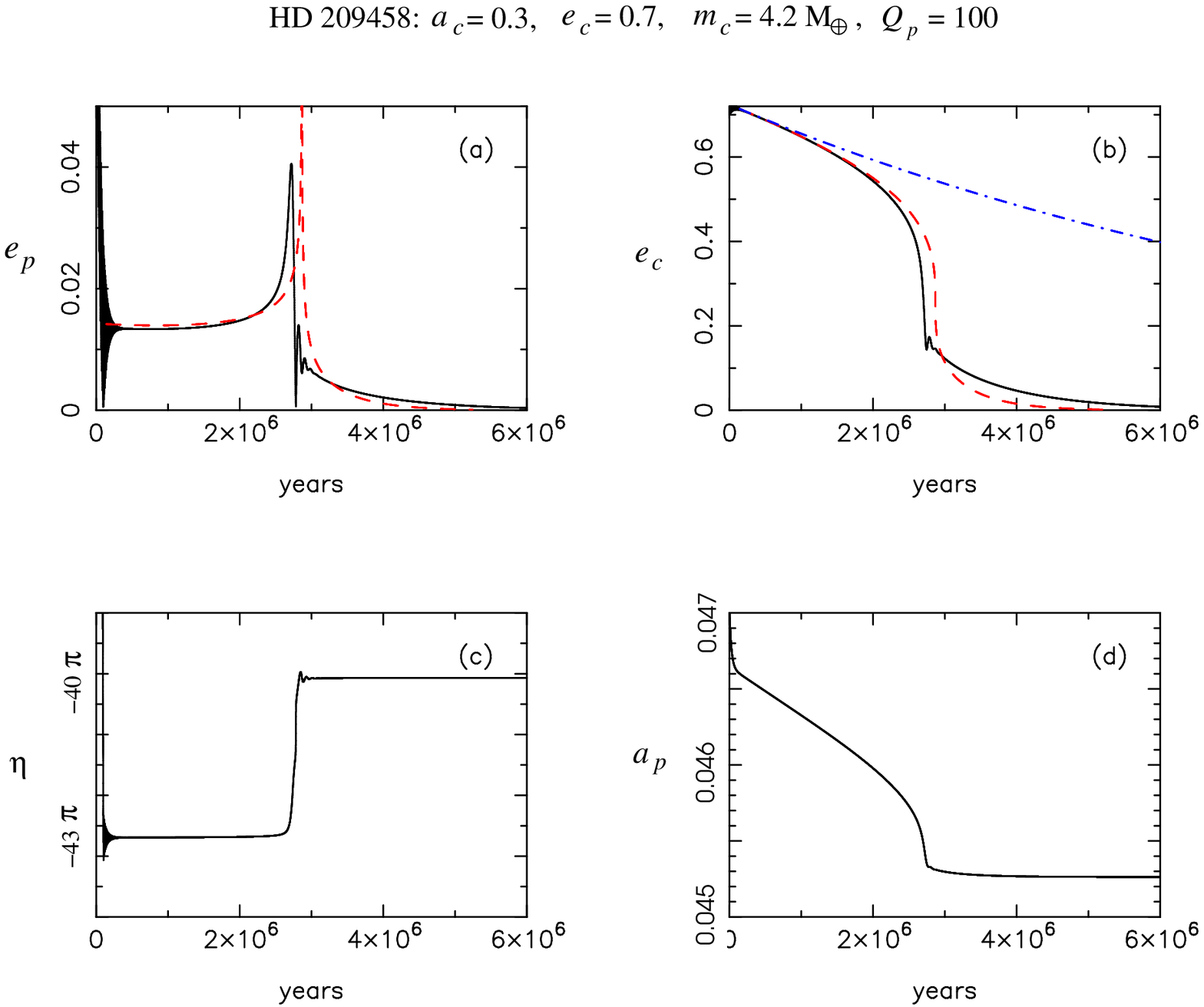}
\caption{A terrestrial-mass planet can sustain the observed eccentricity
of HD 209458 - but not for long enough in this example where $m_c=4.2 M_\oplus$,
at least for $Q_p\lapp2\times 10^5$. {\it Recall that the actual timescale
is proportional to $Q_p$ so that 6 Myr with $Q_p=100$ corresponds
to 6 Gyr with $Q_p=10^5$}. 
The black curves show the numerical solutions while the red dashed curves
are given by the analytical solutions \rn{equilGR} for $e_p$ and
\rn{ecexact} for $e_c$.
(a): The evolution of $e_p$ following capture. After effectively maintaining
the value $e_p^*\simeq 0.013$ for some 2 Gyr (for $Q_p=10^5$),  
$e_p$ increases to a maximum of around 0.04
before rapidly declining to around 0.008 and then slowly declining to zero.
The increase in $e_p$ corresponds to a
singularity in the analytic solution which fairly accurately follows the
full solution.
The fact that the analytical solution is reliable
confirms this dependence.
Thus, for $Q_p=10^5$, the maximum
value of $e_p$ occurs around 2.7 Gyr.
Recall also that the value of $e_p$ at the beginning of the
slow phase is quite insensitive to $e_p(0)$, $\eta(0)$
and $Q_p$.
(b): The evolution of $e_c$ following capture. Again the analytic solution
is a good approximation to the full solution. The blue dot-dashed curve
is the approximate analytic solution given by \rn{ecapprox}. Thus when
a system must pass through the singularity, the timescale given by \rn{tauc}
{\it overestimates} the circularization timescale in contrast to the previous example.
(c): The evolution of $\eta=\varpi_p-\varpi_c$. During the pre-capture
phase, $\eta$ circulates {\it backwards} 
before being captured at $\eta=-43\pi$ (recall that when $\oW<0$,
$\cos\eta\simeq -1$). As the system passes through the singularity, 
that is, as $\oW$ passes through zero to become positive (see 
equation~\rn{oW}), $\eta$ 
circulates back up to $-40\pi$ so that $\cos\eta\simeq 1$. The brief
circulatory period is accompanied by oscillations in $e_p$ and $e_c$
which can be seen in (a) and (b).
(d): The evolution of $a_p$. The full (secular) numerical solution includes
the evolution of both semimajor axes; here we see that over the entire
evolution of this system, $a_p$ changes by less than 4\%. 
}
\label{eup2}
\end{figure}

The solution for the example given in 
Figure~\ref{eta-GR-tides-long} is essentially a modified decaying exponential.
However, there is another regime of the function $f(x;\alpha,\gamma)$
which behaves very differently and is associated with very low-mass
companions which, for a large part of the life of such systems, force
$e_p$ to {\it increase} to some maximum value 
before decaying to zero at the end of the
tidally evolving phase. An example of this is shown in Figure~\ref{eup2}
for the HD 209458 system with a companion of mass
$m_c=0.0125M_J=4.2M_\oplus$ at $a_c=0.3$ AU and an
{\it initial} companion eccentricity of 0.7, a high value chosen
to clearly illustrate the theory. The system enters the slow phase
with $e_c^*=0.705$, $e_p^*=0.013$ and $\eta^*\simeq-43\pi$.
The average reflex velocity of the star due to the companion is $0.64\, \ms$.
See the figure caption for a detailed description of the evolution of the system.
The red dashed curves again represent the analytic solutions \rn{ecexact}
and \rn{equilGR}, clearly demonstrating their predictive power. In fact, the
analytic solution for $e_p$ passes through a singularity as the denominator
in \rn{equilGR} passes through zero. The full numerical solution avoids this,
but still manages to increase to a significant level on a reasonably long
timescale. For $Q_p=10^5$,
passage through the ``singularity'' occurs at 2.2 Gyr, well before the system 
age of 5.5 Gyr, while for $Q_p=2\times10^5$,
passage through this point occurs ``now''.
Note that this system is
stable when the full three-body (non-secular) equations are integrated
including post-Newtonian relativistic terms, tidal and spin
bulges, as well as tidal damping \citep{m1}.

\begin{figure}
\includegraphics[width=150mm]{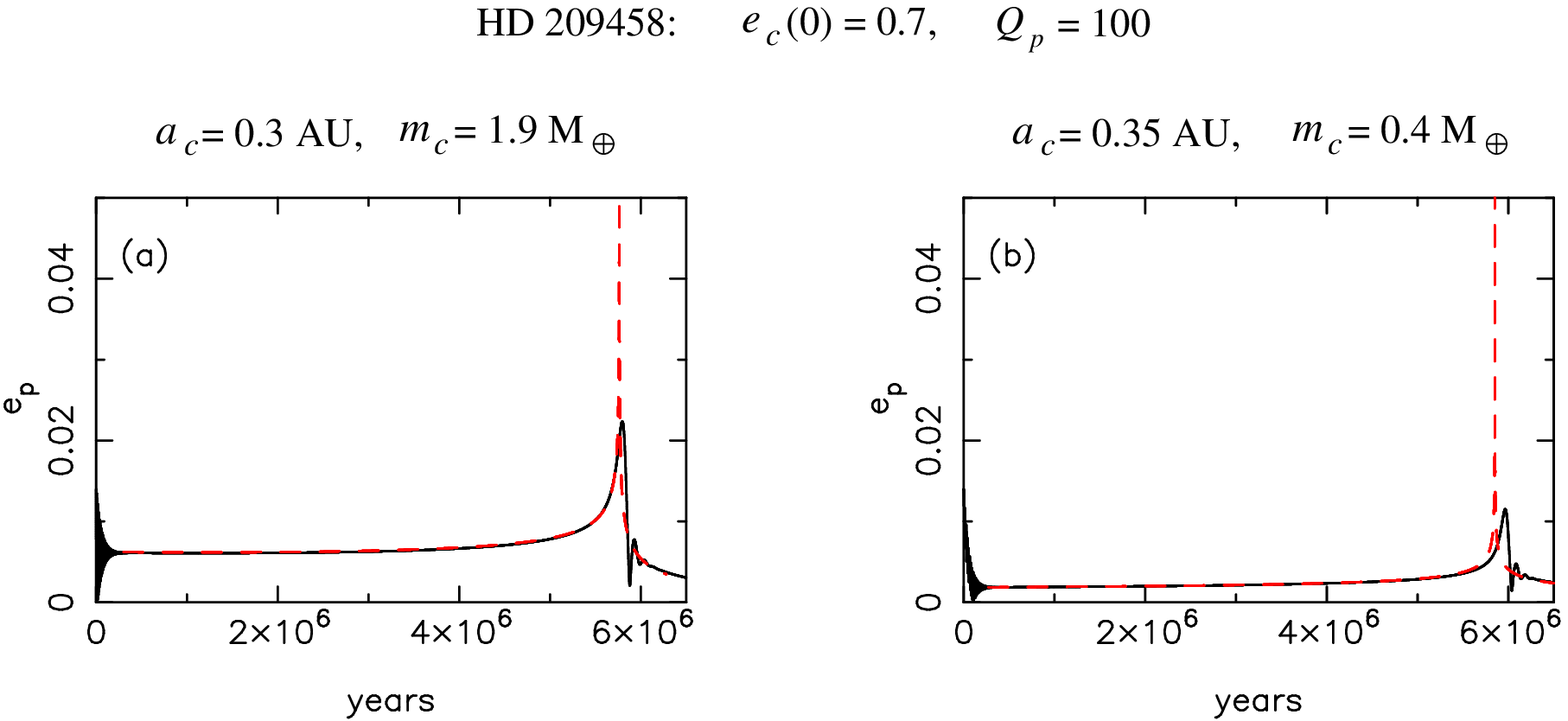}
\caption{Similar to Figure~\ref{eup2} but with lower values for $m_c$.
Passage through $\Delta=0$ occurs later in the life of the system
(compare panel (a) to Figure~\ref{eup2}(a)), while $e_p^*$ is smaller.
For (a), the maximum value of $e_p=0.022$ is reached at
5.8 Gyr for $Q_p=10^5$
(ie., 5.8 Myr for $Q_p=100$), while at 5.5 Gyr, $e_p=0.010$.
For (b), $e_p=0.012$ at 5.96 Gyr, while $e_p=0.004$ at 5.5 Gyr.}
\label{eup3}
\end{figure}

\begin{figure}
\includegraphics[width=170mm]{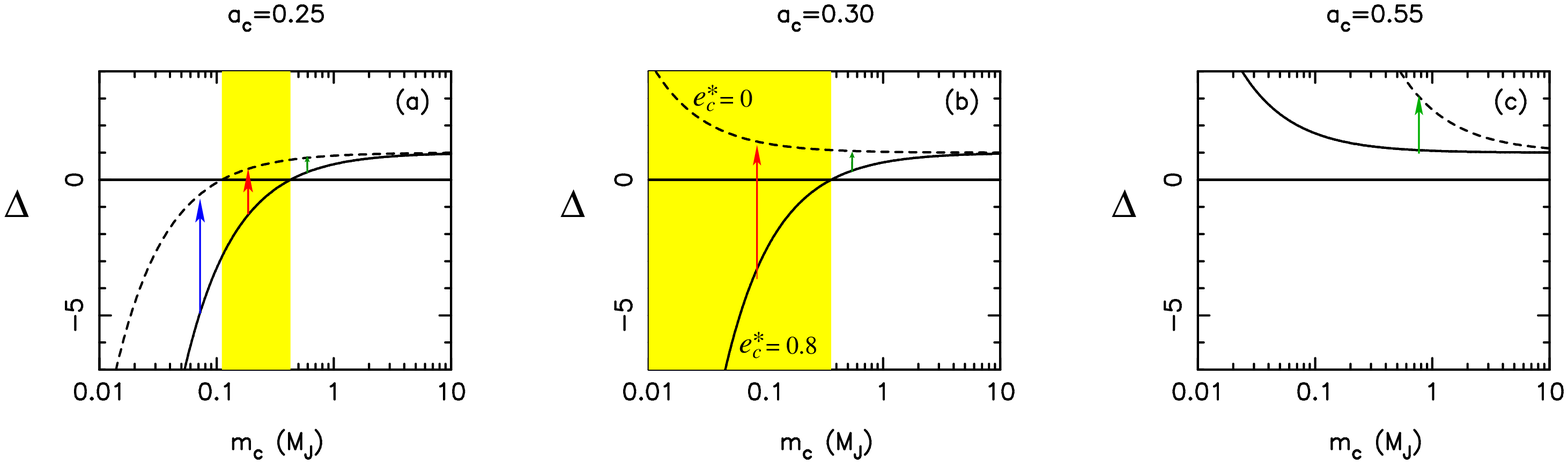}
\caption{Behaviour of the denominator of \rn{equilGR}, $\Delta$,
used to determine which configurations have $e_p^{(eq)}$
necessarily passing through a ``singularity'' during their evolution.
See text for discussion.
Note that some of the systems represented here are not stable when
the full three-body equations are integrated with appropriate damping
\citep{m1}.}
\label{conspire}
\end{figure}

Although systems which pass through the singularity suffer a rapid 
decline in $e_p$ once they have done so, they tend to 
maintain the value of $e_p^*$ that they had at the beginning of the slow phase
before this happens (see also Figure~\ref{eup3}). 
This is in contrast to systems which don't pass
through the singularity; these simply decay approximately 
exponentially. To determine
the parts of parameter space for which $e_p$ necessarily
passes through the singularity and therefore maintains $e_p^*$
before it does so, and in particular which systems pass through the
singularity {\it after}
the current age of the system, we start by studying the behaviour
of the denominator
of \rn{equilGR}. Defining
\be
\Delta(e_c;\alpha,\gamma)=1-\alpha\varepsilon_c^{-1}+\gamma \varepsilon^3,
\ee
where (again), $\alpha=\sqrt{a_p/a_c}(m_p/m_c)$,
Figure~\ref{conspire} plots this function against $m_c$ for specific values of $a_c$
and for $e_c=0.8$ and $e_c=0$. 
There are three regimes of interest. As the system evolves from $e_c^*$
down to $e_c=0$, either
(a) $\Delta$ (and hence $\oW$) remains negative so that $\cos\eta\simeq -1$
always (blue arrow); (b) $\Delta$ passes through zero so that $\eta$ goes 
through a short period of circulation as $\cos\eta$ passes from -1 to 1
(yellow background,
red arrows) or (c) $\Delta$ is positive always and $\cos\eta\simeq 1$ (green arrows).
In the case shown here ($e_c^*=0.8$), systems cross $\Delta=0$ for a
finite range of values of $m_c$ when $a_c\lapp0.29$;
when $0.29\lapp a_c\lapp 0.52$, all systems with $m_c$ less than 
some critical value cross
$\Delta=0$; when $a_c\gapp 0.52$ {\it no} systems cross $\Delta=0$.
Figure~\ref{conspire}(b), for example, 
shows that all systems with $m_c\lapp 0.25 M_J$
and $a_c=0.3$ (and $e_c^*=0.8$)
necessarily evolve through $\Delta=0$.
The ranges above vary with different values of $e_c^*$;
curves corresponding to $e_c^*<0.8$ lie between the two shown.
What remains to be determined, however, is which of these systems
maintain a non-zero value of $e_p$ for at least as long as the current age of the
system before they do so. Once this is done, the parameter ranges should be
further restricted by determining which systems are stable \citep{m7}.

\begin{figure}
\includegraphics[width=150mm]{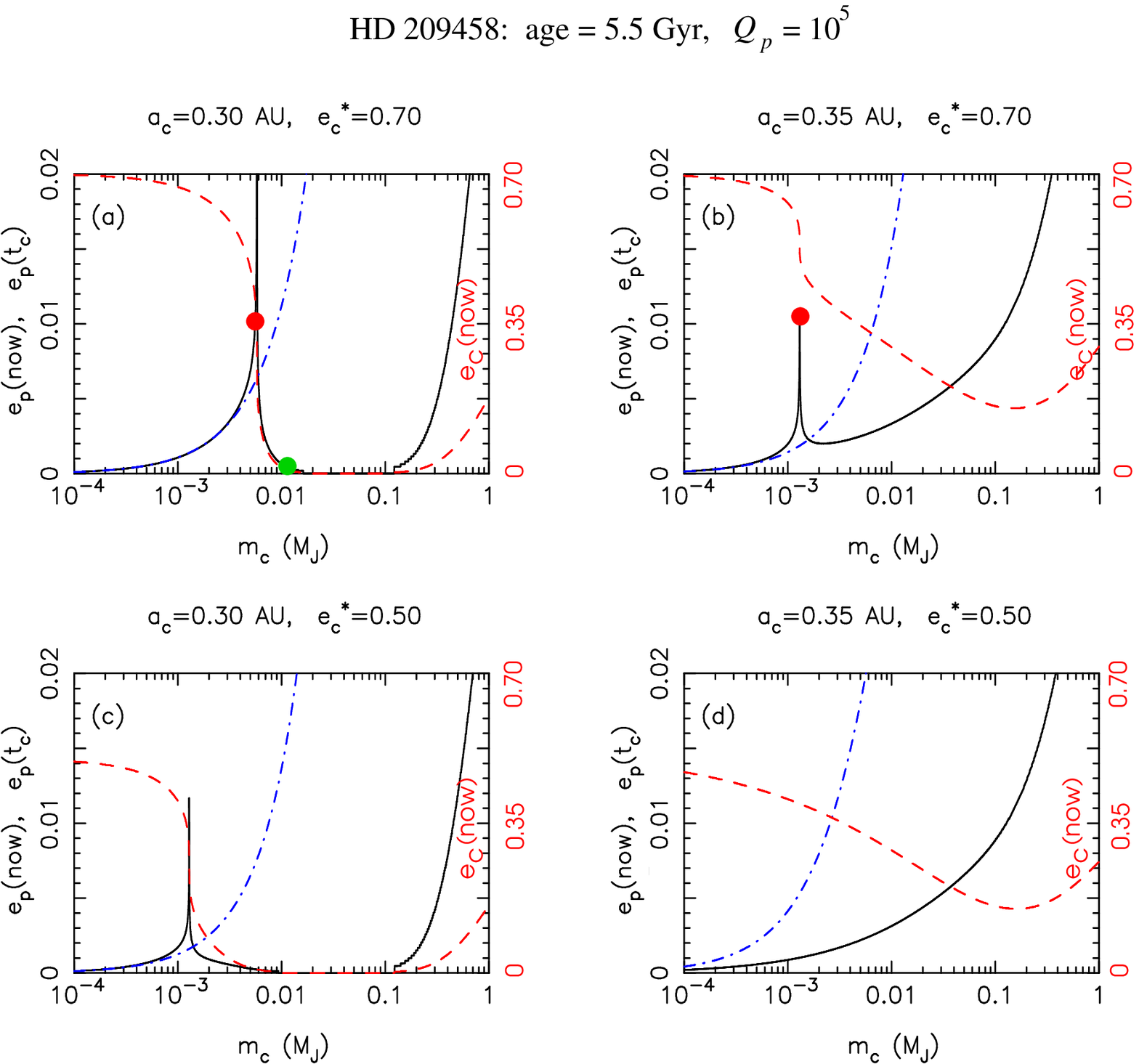}
\caption{
The values of $e_p$ and $e_c$ (black solid and
red-dash curves respectively) at $t=5.5$ Gyr, and the value of $e_p$
at the beginning of the slow phase ($e_p(t_c)$: blue dot-dash curves), all with $Q_p=10^5$. From Figures~\ref{eup2} 
and \ref{eup3} it is
evident that systems which pass through $\Delta=0$ in their lifetime
approximately maintain their value of $e_p$ at $t=t_c$ before they do so. 
Thus, for example, 
in (a) all systems with $m_c\lapp 0.004 M_J$ will have the same
finite (but small) eccentricities they had when they entered the slow phase.
For a small range of masses ($0.004<m_c/M_J<0.006$) $e_p({\rm now})>e_p^*$
(the black curve sits above the blue curve), while for greater masses
$e_p({\rm now})\ll e_p^*$.
Note that formally these curves have no upper bound
at the point corresponding to $\Delta=0$, however the finite 
resolution of the plots
suggest erroneously that they do. In reality there
does exist an upper bound when the secular equations
are directly integrated
as is evident in Figures~\ref{eup2}
and \ref{eup3}. It is clear from this figure that high initial values
of $e_c^*$ are required to have very low-mass companions
responsible for maintaining the eccentricity of inflated planets,
although lower values of $e_c^*$ are required for lower values
of $a_c$ 
(see Figure~\ref{HD209458d} and 
Section~\ref{discussion} for a discussion on this point).
}
\label{epec}
\end{figure}

Figure \ref{epec}
plots $e_p({\rm now})$,
$e_p(t_c)$ and $e_c({\rm now})$ against companion mass
for the HD 209458 system for various values of $a_c$ and $e_c^*$.
Here ``now'' refers to $t=5.5$ Gyr, and the quantities are
calculated using \rn{equilGR} and \rn{ecexact}. The green-filled circle
corresponds to the system shown in Figure~\ref{eup2} at
$t=5.5$ Gyr for $Q_p=10^5$ ($t=5.5$ Myr for $Q_p=10^2$), while
the red-filled circles correspond to Figure~\ref{eup3} at the same time.
Note the extremely low companion masses in Figures~\ref{eup3}(a)
and (b): $m_c=1.9M_\oplus$ and $m_c=0.4M_\oplus$ respectively.
The maximum values of $e_p$ achieved are (a) 0.022 at 5.8 Gyr
and (b) 0.012 at 6.0 Gyr.
The corresponding values of $e_c$ are 0.27 and 0.50.
At 5.5 Gyr, the values for $e_p$ are (a) 0.010 and (b) 0.004, with
corresponding values for $e_c$ of 0.41 and 0.60.
Thus scenario (a) represents a possible solution for HD 209458 for 
a value of $Q_p$ around $10^5$, while scenario (b)
requires a $Q$-value
around 10\% smaller in order that the maximum value of $e_p$
occurs about now (recall that this also depends on the age
estimate for the star).
Figure~\ref{epec} demonstrates that high initial values
of $e_c^*$ are required in order that extremely low-mass companions
be responsible for maintaining the eccentricity of inflated planets
(see Section~\ref{discussion} for a discussion on this point).

\begin{figure}
\includegraphics[width=150mm]{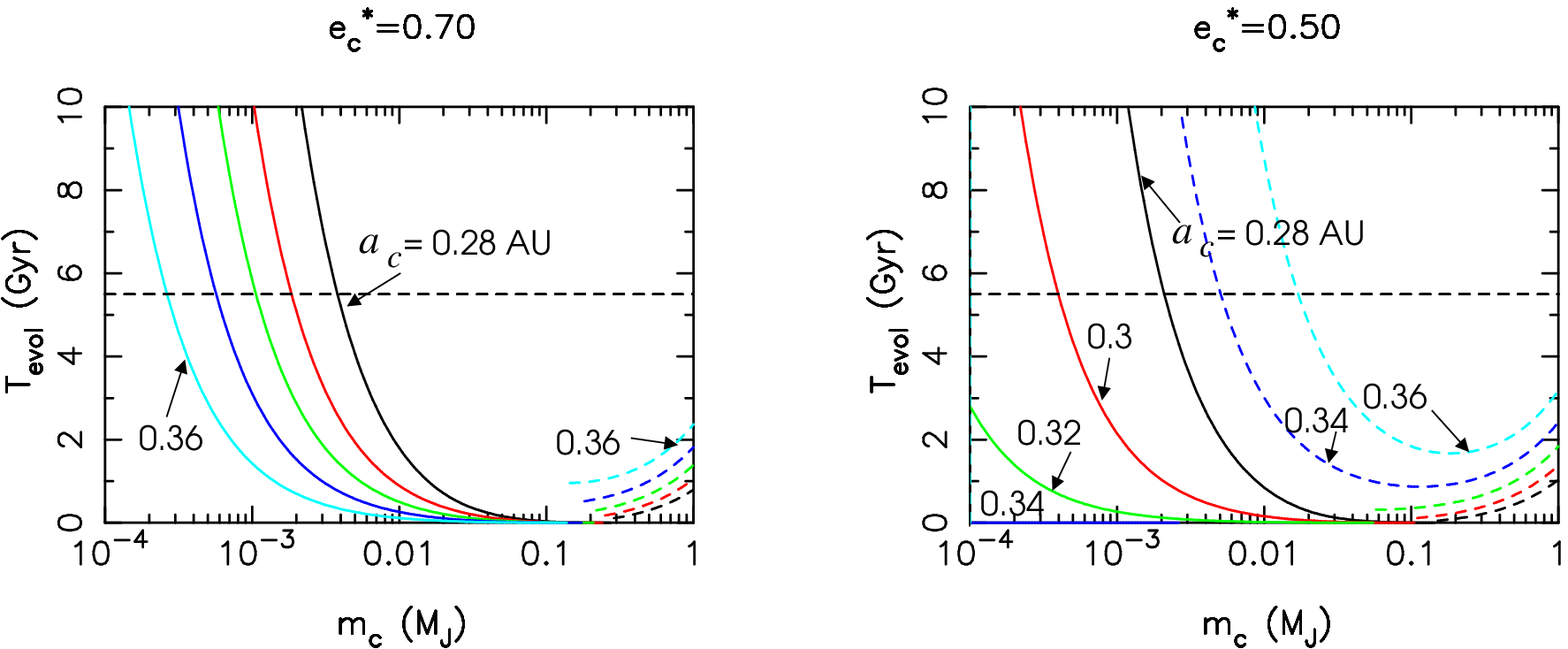}
\caption{The time to reach the maximum value of $e_p$, $T_{evol}$, for
systems which pass through $\Delta=0$ (solid curves), and
the evolution timescale (defined as the time
corresponding to $e_c=e_c^*/e$) for systems which do not (dashed curves).
Various values of $a_c$ are indicated on the plots;
the estimated age of HD 209458 is indicated by the horizontal dashed line.
The lower the companion mass, the longer the system takes before
it passes through $\Delta=0$ (given that it does so at all).
This phenomenon favours systems with high values of $e_c^*$.}
\label{Tevolve}
\end{figure}

\begin{figure}
\includegraphics[width=150mm]{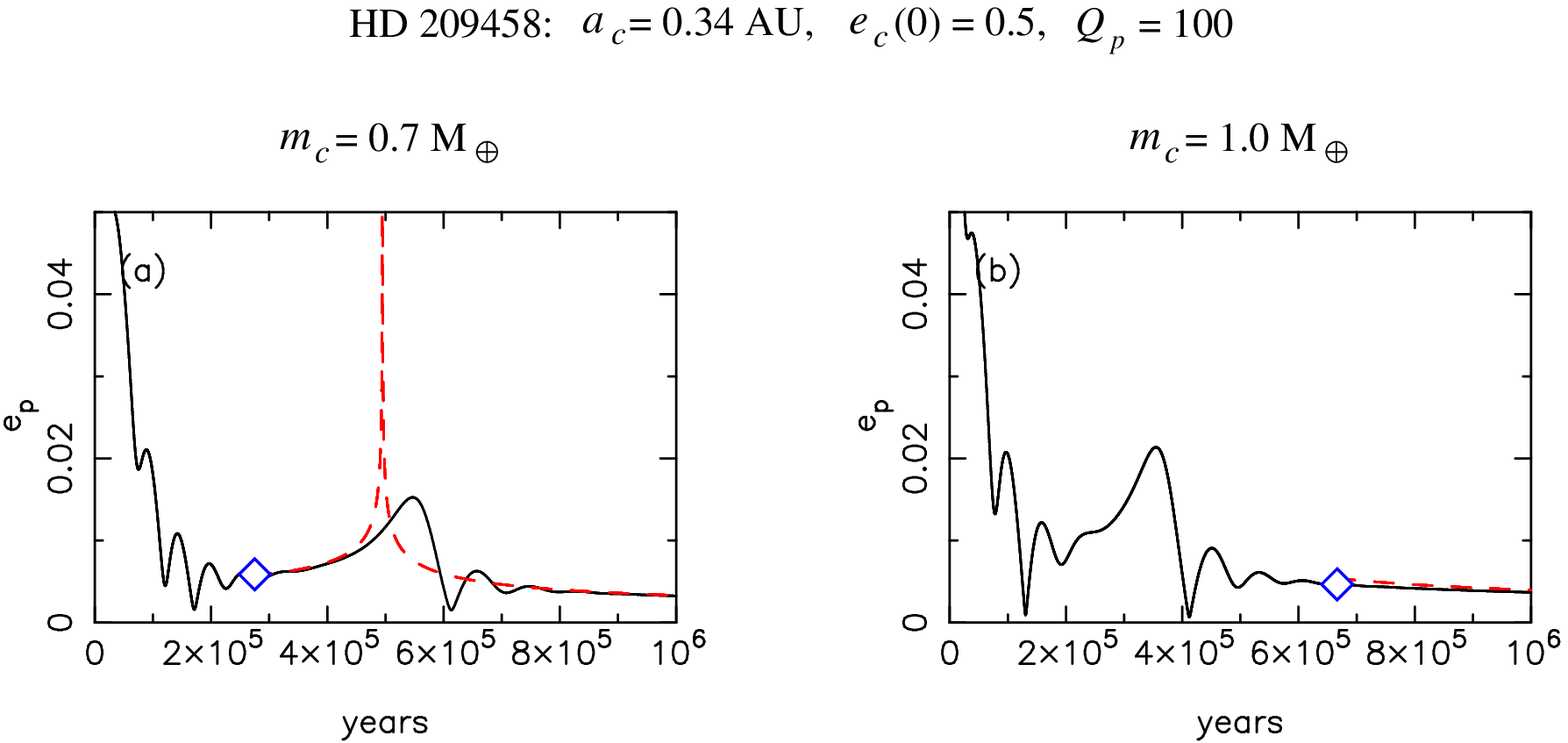}
\caption{Demonstration that for some systems, passage through
$\Delta=0$ occurs very early on in the evolution, and circularization
is approximately exponential after that. The blue diamond symbols
again indicate the formal beginning of the slow phase.}
\label{eup4}
\end{figure}

Figure \ref{epec}, together with Figures~\ref{eup2} and \ref{eup3}
and equation~\rn{tautrue},
suggest that for given values of $a_c$ and $e_c^*$,
the time a system takes to achieve its maximum value of $e_p$ (given
it passes through $\Delta=0$)
increases as the companion mass decreases. 
Figure~\ref{Tevolve}
plots this time, $T_{\rm evol}$, as a function of $m_c$ for various
values of $a_c$ and $e_c^*$ (solid curves). Also plotted is
$\tau_c^{true}$ (equation~\rn{tautrue}) which is 
the $e$-folding time for systems which don't go through $\Delta=0$
but simply decay approximately 
exponentially (dashed curves). 
The small values of $T_{\rm evol}$ corresponding to,
for example, systems for which $e_c^*=0.5$ and $a_c=0.34$ with
$m_c\lapp 0.002M_J$ (the dark blue curve which appears
coincident with the $x$-axis), compared with the large values of $T_{\rm evol}$
for $m_c\gapp 0.003M_J$,
simply indicate that passage through
$\Delta=0$ occurs early in the life of the system, while
decay following passage is very slow. Figure~\ref{eup4}
illustrates this with the above example. The blue diamond indicates
the nominal beginning of the slow phase (as defined in the discussion
following equation~\rn{tauc}). There is little difference between
Figures~\ref{eup4}(a) and (b) except that \ref{eup4}(b) does not
meet the slow phase criterion until after the system passes through
$\Delta=0$.

It is of interest to ask whether or not it is possible for a system to skip
the librating phase
and evolve directly to the true fixed point $(e_p,e_c)=(0,0)$.
Referring to equations~\rn{lib0r} and \rn{libpir} we see that for 
$e_p<2e_p^{(eq)}$ a system will librate as long as 
$-\pi/2<\eta<\pi/2$ or $\pi/2<\eta<3\pi/2$. Since a circulating system
must pass through all values of $\eta$, once $e_p$ is permanently below
$2e_p^{(eq)}$ it has no choice but to librate until a local quasi-fixed point
is reached.

\section{Constraints on orbital parameters of companions}
We are now in the position to constrain the orbital parameters and
masses of possible companion planets to HD 209458b,
HAT-P1b and WASP-1b. 
Given estimated ages,
our aim is to find initial conditions and companion
masses for which these planets have eccentricities in the
ranges observed (WASP-1b is currently unconstrained), and
in the case of HD 209458, produces a stellar reflex velocity less than
$10\,\ms$ (an upper limit which gives companion masses less than
the upper bound of $0.3 M_J$ cited in \citet{l1}).

Acknowledging again that these systems are unlikely to be exactly 
coplanar (but see Figure~\ref{relinc} which shows that moderate
inclination has little effect), 
we assume zero mutual inclination and vary the parameters
$a_c$,  $m_c$ and $e_c(0)$.
Our results correspond to the case $e_p(0)=0.1$ and $\eta=170^o$,
but are not very sensitive to using other values of these parameters
(ie, they evolve to the 
quasi-fixed point $(e_p^*,e_c^*,\eta^*)$ initially with a variation
in $e_c^*$ of at most around 10\%).
Note also that while our aim is to determine whether or not current-day solutions exist
(especially for HD 209458 which is the most constrained),
we can also say something about the orbital elements at earlier times
and hence (cautiously) constrain the initial conditions.

\subsection{HD 209458}

\begin{figure}
\includegraphics[width=150mm]{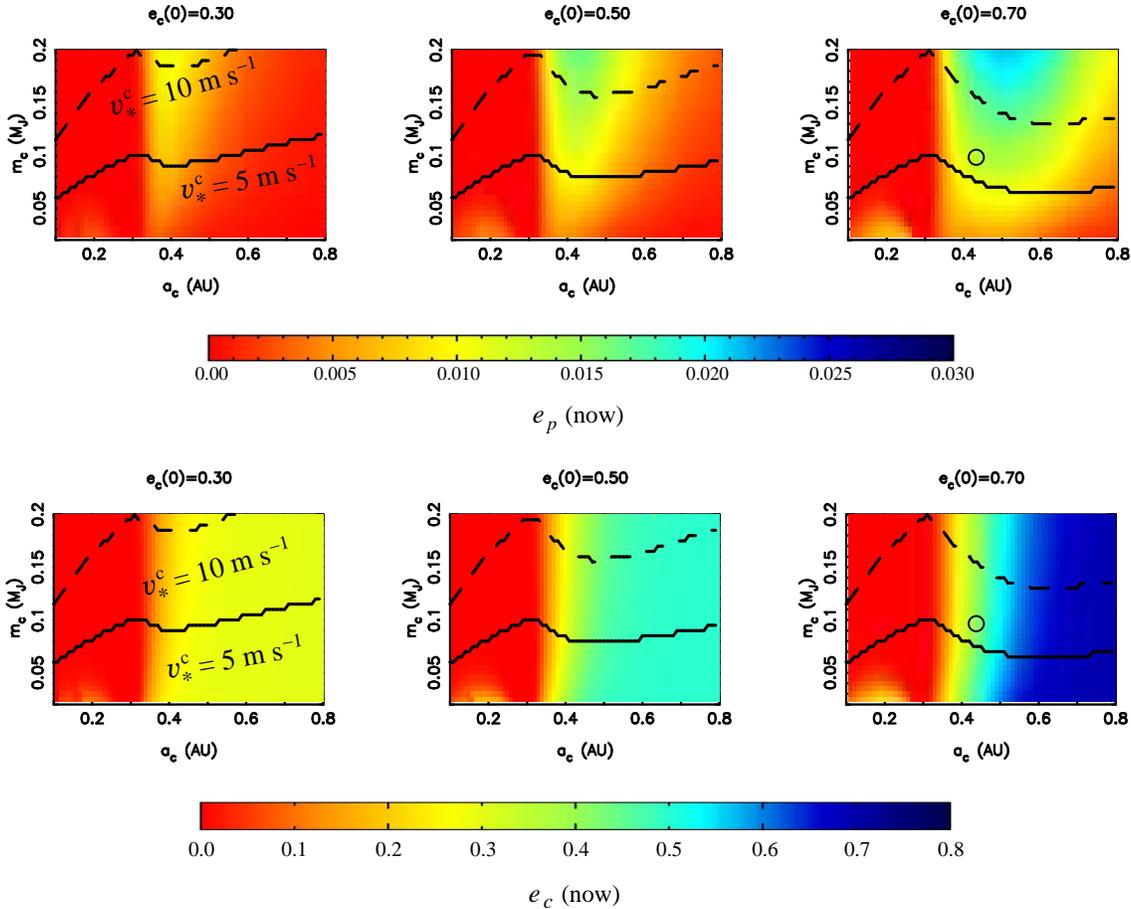}
\caption{``Current'' values of the eccentricity of HD 209458b, $e_p$(now), (top three panels)
and of its hypothetical companion, $e_c$(now), (bottom three panels) 
for a range of hypothetical companion masses and semimajor axes.
An estimated age for the system of 5.5 Gyr is assumed 
as well as $Q_p=10^5$. Initial values for
$e_c$ are indicated at the top of each panel, while $e_p(0)=0.1$ 
and $\eta(0)=170^o$ for
all experiments. 
The minimum companion mass is $m_c=0.02M_J=6.67M_\oplus$.
Note the circle on the panels corresponding to $e_c(0)=0.7$: although
the companion orbit starts with a high eccentricity, its current eccentricity
is relatively low (around 0.35).
Also shown are curves of constant stellar reflex velocity, $v_*^c$, due to
the companion {\it at periastron}.
See text for discussion.
}
\label{HD209458b}
\end{figure}

\begin{figure}
\includegraphics[width=150mm]{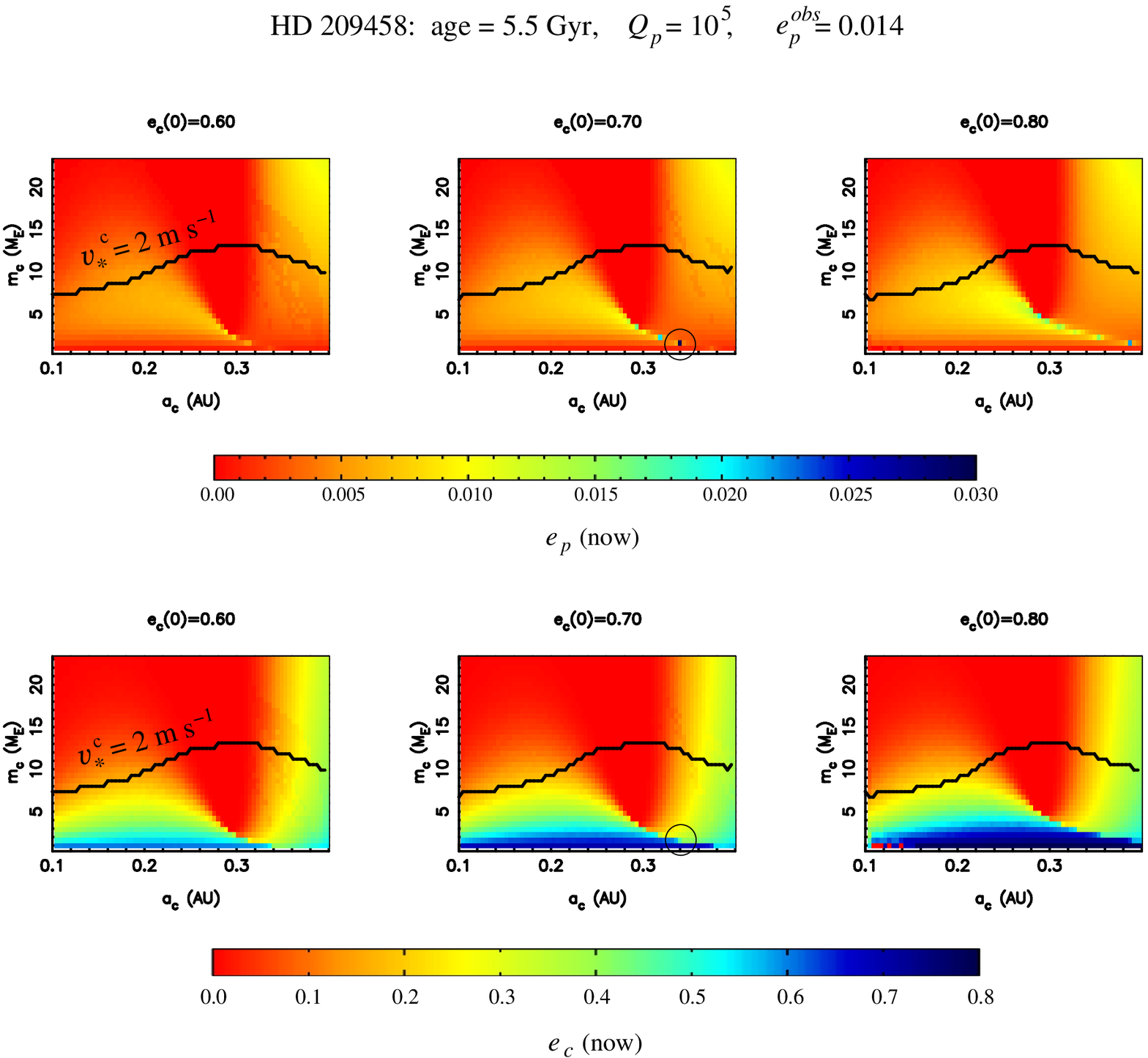}
\caption{
Detail of the lower left-hand corners of two of the panels
in Figure~\ref{HD209458b} ($e_c(0)=0.70$) together with
those for $e_c(0)=0.60$ and $e_c(0)=0.80$.
These indicate that ``solutions'' exist
even for companion masses as low as an Earth mass,
and that the number of solutions increases with increasing $e_c(0)$.
Note in particular the blue dots in the top panels indicating
relatively high eccentricities: these correspond to the maxima
see in Figures~\ref{eup2} and \ref{eup3}. 
The solution circled corresponds to $m_c\simeq 1.0M_\oplus$
with $e_p({\rm now})=0.03$, $e_c({\rm now})\simeq 0.5$ and $a_c=0.34$ AU.
Note also that not all systems represented here would be stable
had a direct integration code been used to do the integrations \citep{m4}.
The minimum companion mass is $m_c=0.001M_J=0.33M_\oplus$.
Also shown are curves of constant stellar reflex velocity, $v_*^c$, due to
the companion at periastron.
}
\label{HD209458c}
\end{figure}

\begin{figure}
\includegraphics[width=145mm]{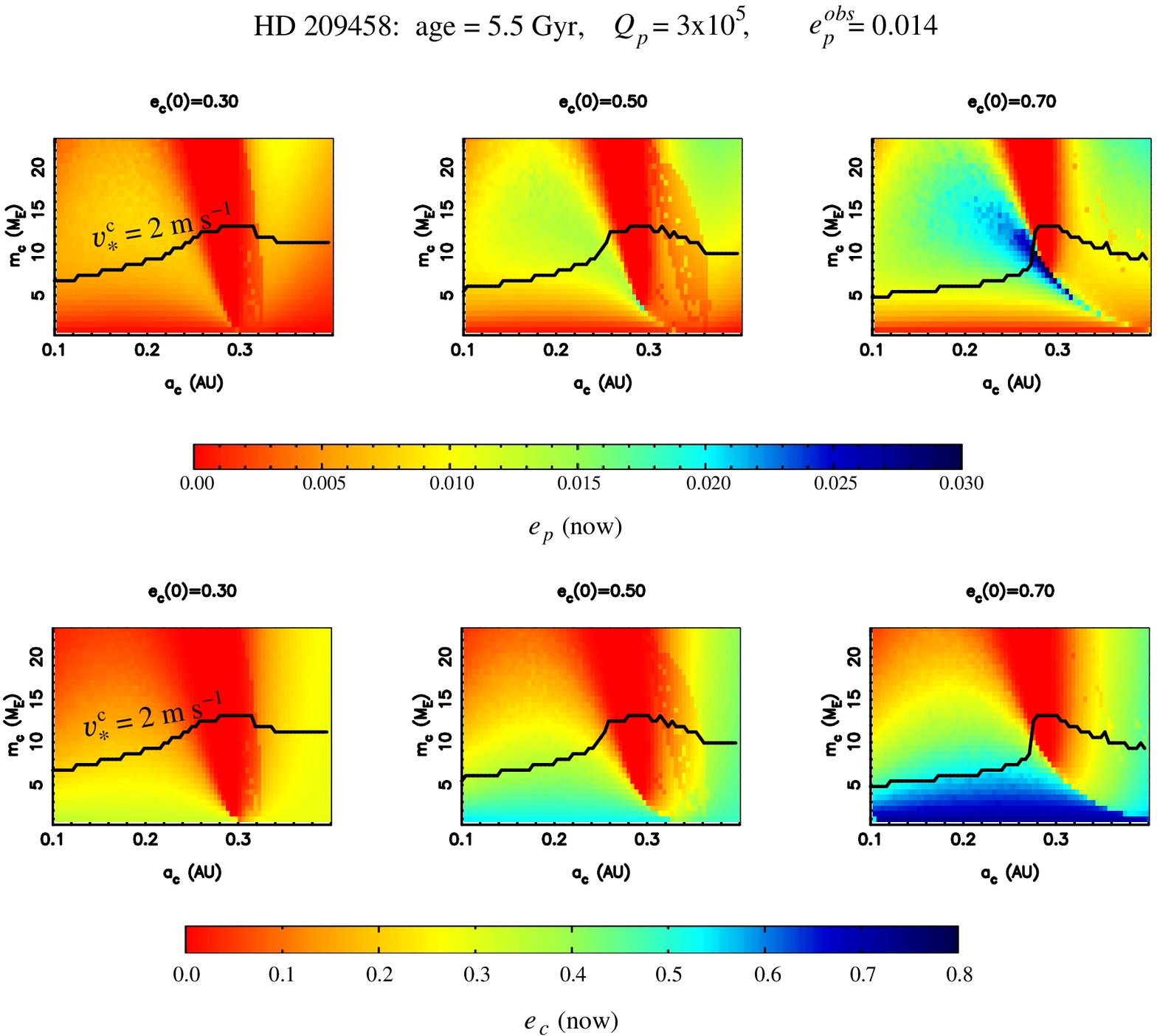}
\caption{
Similar to Figure~\ref{HD209458c} but with $Q_p=3\times 10^5$
showing the dramatic effect increasing $Q_p$ has on $e_p({\rm now})$.
Note the general reduction in the values of $e_c$ from $e_c(0)$
to $e_c$(now).
Solutions exist for $e_c(0)\gapp 0.4$, an initial eccentricity not 
inconsistent with those found by \citet{m6}; higher values of $e_c(0)$
yield more solutions.
Also shown are curves of constant stellar reflex velocity, $v_*^c$,  due to
the companion at periastron.
}
\label{HD209458d}
\end{figure}

Figure~\ref{HD209458b}
presents results for $\eta(0)=170^o$, $e_p(0)=0.1$, $0.3\leq e_c(0)\leq 0.8$,
$0.02\leq m_c/M_J\leq 0.2$ and $0.1\leq a_c/{\rm AU}\leq 0.8$.
The full secular equations \citep{m2} were integrated with $Q_p=100$ 
until $e_c^*$
was reached and the value for this was used in \rn{ecexact} to calculate
the current
value of $e_c$ (``$e_c$(now)" )
and hence of $e_p$ (``$e_p$(now)'') via \rn{equilGR}
for a given {\it actual} $Q$-value
($Q_p=10^5$).
Recall from the discussion in
Section~\ref{dissipation} that timescales scale linearly with $Q_p$ and that
{\it the relative change in $e_c^*$ is less than 2\% over 
four orders of magnitude in $Q_p$}.
Each point in the $m_c-a_c$ plane is coloured according to 
$e_p(now)$ (top six panels) and $e_c(now)$ (bottom six panels).
Also shown are boundaries for stellar reflex velocities {\it at periastron} 
$v_*^c=10\,\ms$ and
$v_*^c=5\,\ms$. 
Defining a ``solution'' to be a configuration for which 
$e_p({\rm now})>0.010$,
solutions exist below  $v_*^c=10\,\ms$ for
$e_c(0)\geq 0.3$, $e_c({\rm now})\gapp 0.2$, $a_c\gapp 0.35$ AU 
and $m_c\gapp 0.05\, M_J$,
and below
$v_*^c=5\,\ms$ for
$e_c(0)\geq 0.5$, $e_c({\rm now})\gapp 0.3$, $a_c\gapp 0.4$ AU 
and $m_c\gapp 0.05\, M_J$.
However, notice the orange patches around $a_c$=0.2 AU corresponding
to $m_c<0.03\,M_J$: enlargements of these are shown in 
Figure~\ref{HD209458c}.
The slow phase of the evolution of these systems actually
involves an {\it increase} in $e_p$ as discussed in Section~\ref{incep}, 
so that companion
masses as low as $1M_\oplus$ are capable of exciting significant
eccentricity in the observed planet. Some systems reach
eccentricities around 0.02 (the aqua points in the top three
plots of Figure~\ref{HD209458c}). This effect is more pronounced
when $Q_p$ is higher as is demonstrated in
Figure~\ref{HD209458d}
for the case $Q=3\times 10^5$.

\subsection{HAT-P1}\label{hat}

HAT-P1 is a member of a binary system for which the 
projected separation of the stars is 1550 AU \citep{b1}.
At this distance, the companion is incapable of inducing any
eccentricity in the planet's orbit because the tidal damping timescale
is far shorter than any timescale on which eccentricity could be induced.

\begin{figure}
\includegraphics[width=150mm]{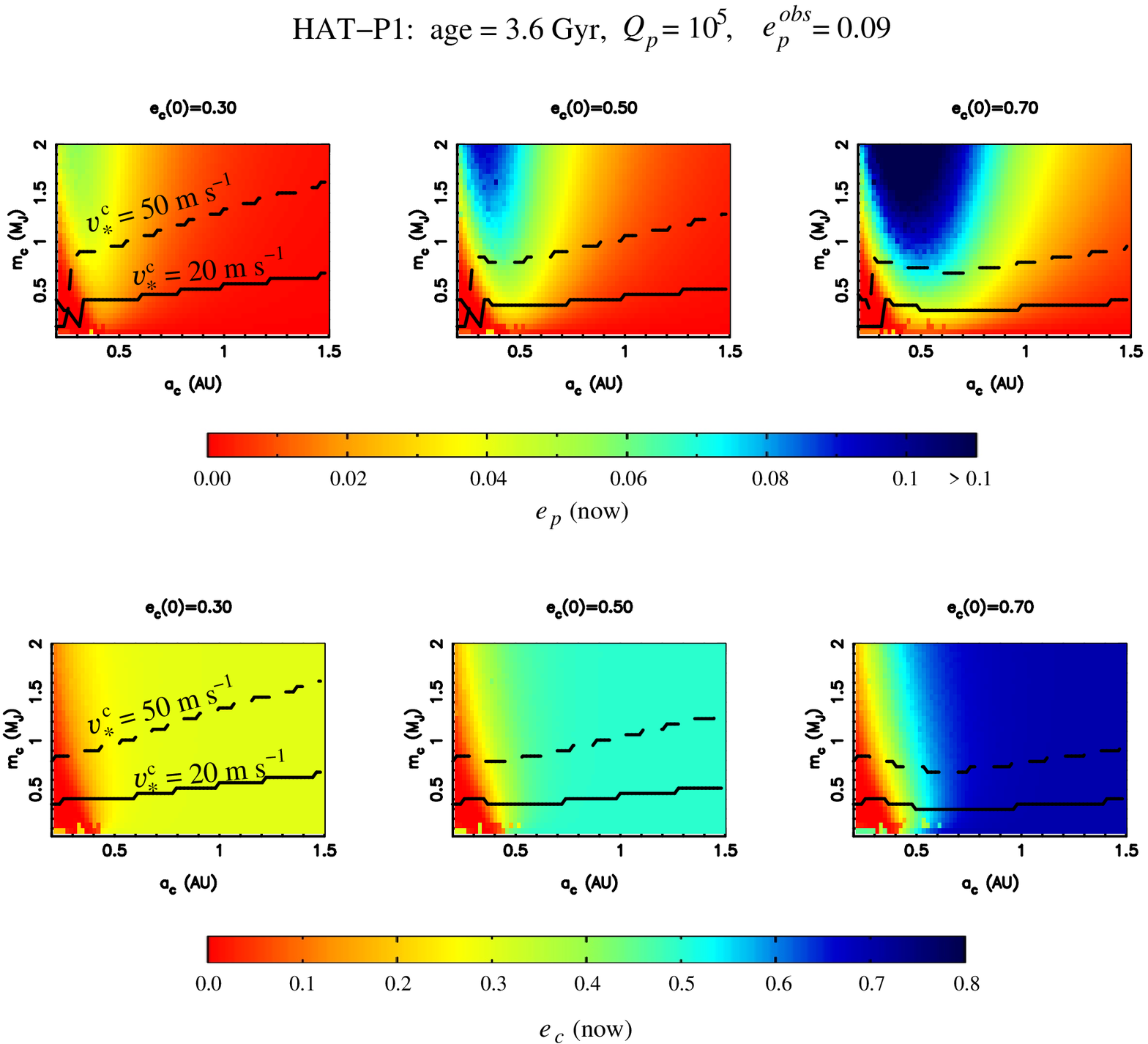}
\caption{``Current'' values of the eccentricity of HAT-P1b, $e_p$(now), (top three panels)
and of its hypothetical companion, $e_c$(now), (bottom three panels) 
for a range of hypothetical companion masses and semimajor axes.
An estimated age for the system of 3.6 Gyr is assumed 
as well as $Q_p=10^5$. Initial values for
$e_c$ are indicated at the top of each panel, while $e_p(0)=0.1$ 
and $\eta(0)=170^o$ for
all experiments. 
 Initial values for
$e_c$ are indicated at the top of each panel, while $e_p(0)=0.1$ 
and $\eta(0)=170^o$ for
all experiments. 
Note the scale used for $e_p({\rm now})$
is different to that used for HD 209458.
The tentative observed
eccentricity is $0.09\pm0.02$. 
Also shown are curves of constant stellar reflex velocity, $v_*^c$,  due to
the companion at periastron.
See text for discussion.
}
\label{HATb}
\end{figure}

\begin{figure}
\includegraphics[width=170mm]{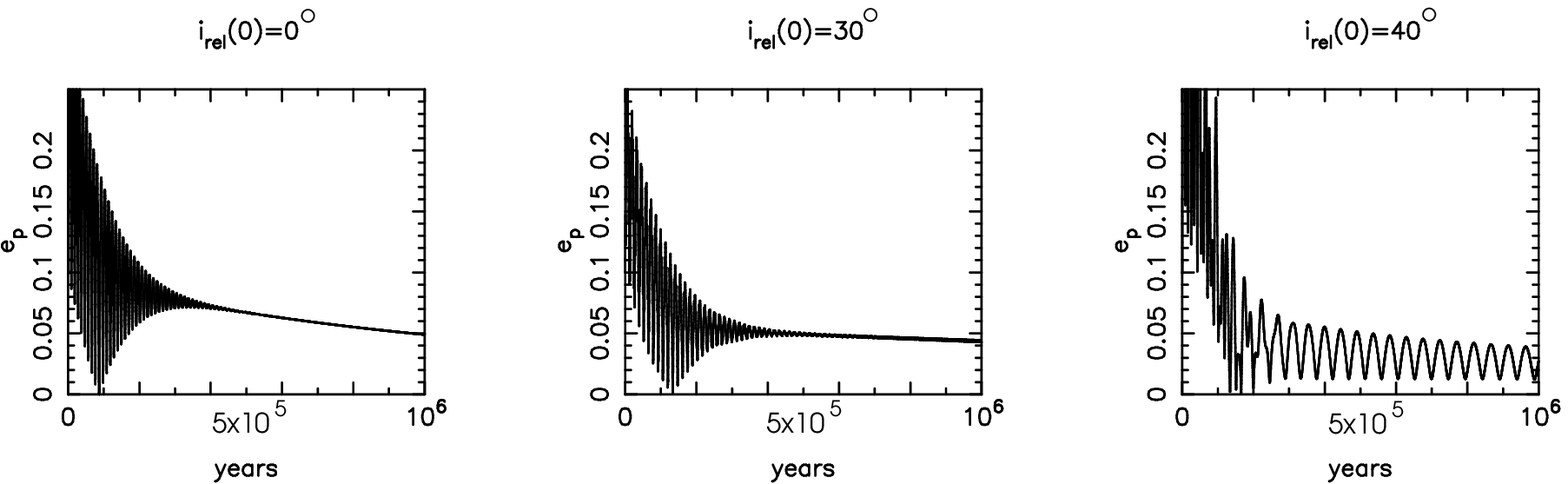}
\caption{The effect of inclination. When the orbits are mutually
inclined the value of $e_c^*$ is reduced slightly, and the eccentricity
is no longer non-oscillatory during the slow phase, although the oscillation
amplitude is small for moderate mutual inclinations ($\lapp 35^o$).
Kozai cycles are present here for $i_{rel}=40^o$.
This example is for a HAT-P1-type system with 
$a_c(0)=0.4$, $e_c(0)=0.5$, $m_c=0.5M_J$
and $Q_p$ again set artificially low at 100.}
\label{relinc}
\end{figure}

\begin{figure}
\includegraphics[width=140mm]{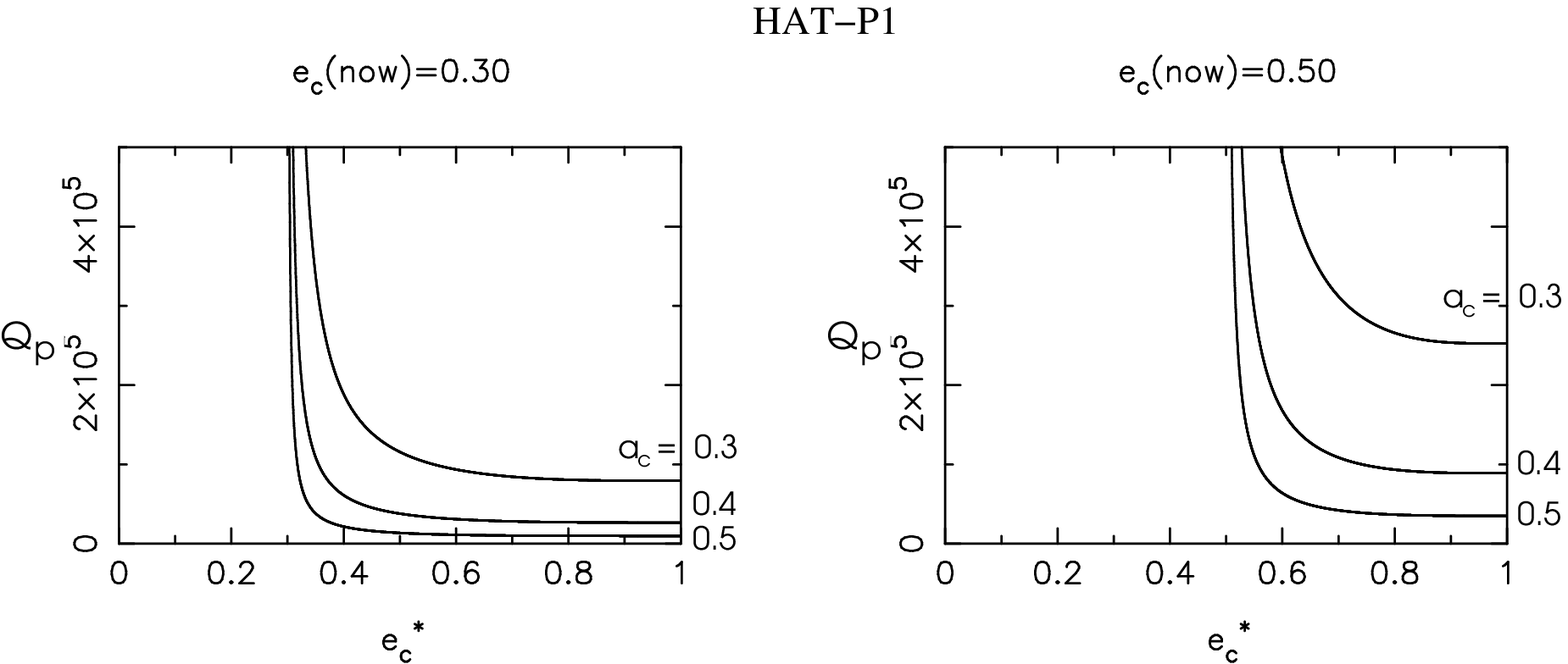}
\caption{Using observed data to put a lower bound on $Q_p$ for 
HAT-P1b. Example shown here is for $m_c=1M_J$. See text for discussion.
}
\label{Qpec}
\end{figure}

As far as the author is aware, at the time of writing 
no upper bound has been put on the reflex velocity of the star
due to a companion planet. The preliminary value of $e_p=0.09\pm 0.02$
awaits refinement, however, as \citet{b1} note, it produces a similar
tidal dissipation rate in the planet to HD 209458b, consistent with its
similar radius and
its increased distance from the star (see Table~\ref{theory}).
We have performed similar experiments for this system as
was done for HD 209458, 
except for a wider range of companion masses and semimajor axes.
The results of these are presented in
Figure~\ref{HATb}
for $Q_p=10^5$.
These results suggest that if a companion planet is responsible for
the inflated radius of HAT-P1b, it should be detectable in the
radial velocity signal, except if a very low-mass companion is
responsible, solutions for which are similar to those for HD 209458
(not visible in Figure~\ref{HATb}).
If a companion planet is detected and a reliable measurement
of the eccentricity of HAT-P1b is made (as well as reliable estimates
for the companion's mass, 
semimajor axis and eccentricity), it may be possible to use the 
theory developed here to deduce a relationship between the $Q$-value of HAT-P1b
and $e_c^*$, and moreover,
to put a lower bound on the value of $Q_p$ (recall that $e_c^*$ is the companion's
eccentricity when the system enters the slow phase and hence is an
an estimate of the initial eccentricity). 
The procedure is as follows.
\ben
\item
Check how accurately equation \rn{equilGR} giving the relationship between $e_p$ and 
$a_p$, $a_c$, $m_p$, $m_c$ and $e_c$ holds.
Any substantial departure will be due to either
(a) a nearby third planet; or (b)
a significantly non-zero mutual inclination; or (c)
a $Q$-value which is large enough for the system not to have yet
entered the slow phase. 
In case (a), a third planet will modulate the eccentricity
of its inner siblings, thereby preventing the equilibriating process
from occurring (unless $3\tau_c^{true}$ is less than the age
of the system, in which
case the the process can work on all three planets!)
In case (b), the effect of mutual inclination is shown in Figure~\ref{relinc}.
The value of $e_c^*$ is slightly reduced, and
the eccentricity
is no longer non-oscillatory during the slow phase, although the oscillation
amplitude is small for moderate mutual inclinations.
The mutual inclination itself oscillates with an amplitude
of around $20^o$. Inclinations need to be greater than around
$40^o$ to signficantly affect the evolution.
In case (c), since
$t_c\simeq 3\,\tau_{circ}\simeq 0.33$ Gyr for $Q_p=10^5$,
$Q_p$ would  have to be greater than $10^6$ to
still be substantially far from pseudo-equilibrium (given an age of 3.6 Gyr).
\item
Given that \rn{equilGR} is satisfied, 
\rn{ecaccurate} can be used with $t=3.6$ Gyr and $t_c=3\,\tau_{circ}$ 
to give a relationship between $Q_p$ and $e_c^*$
and in particular, to put a lower bound on $Q_p$.
Examples of this are plotted
in Figure~\ref{Qpec} for $m_c=1M_J$ and for several values of $a_c$ and 
$e_c$(now). For example, if a companion is found at
$a_c=0.4$AU with $e_c=0.5$ (corresponding to $e_p=0.11$), one can conclude that $Q_p>1.0\times 10^5$.
\een

\subsection{WASP-1}

While \citet{b2} find that around four times as much energy is required
to inflate WASP-1b to its observed size than
for HD 209458b (they used a value for its mass of $0.87M_J$;
its revised estimate is $0.79M_J$ compared
to $0.64M_J$ for HD 209458), its closer proximity to the parent star
results in a similar value for $e_p$ for both systems (see Table~\ref{planets}).
While there is currently no published estimate for the eccentricity
of WASP-1b, numerical experiments performed for this system with
an estimated age of 2 Gyr
show a range of solutions very similar to those for HAT-P1 (Figure~\ref{HATb})
{\it but with the legend of values for $e_p$(now) replaced by one with
a maximum of 0.05 instead of 0.1}.
Very low-mass solutions also exist for this system, albeit for a smaller
range of parameters (or higher $Q$-values) than for HD 209458b and
HAT-P1b.
Note that
Table~\ref{theory} suggests $e_p$ is likely to be similar to that of HD 209458
if a companion is responsible for its inflated radius.

\begin{figure}
\includegraphics[width=140mm]{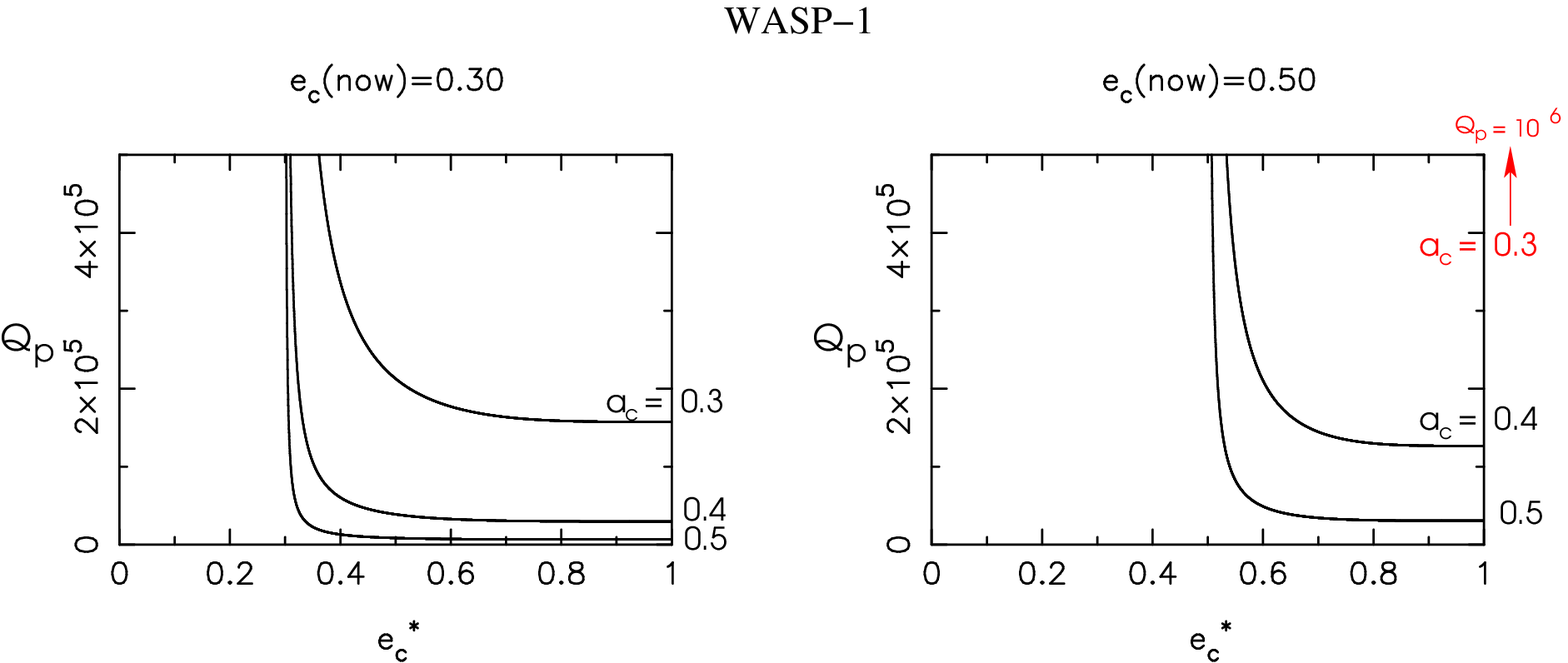}
\caption{Using observed data to put a lower bound on $Q_p$ for 
WASP-1b. Example shown here is for $m_c=1M_J$. See text for discussion.
}
\label{QpecWASP}
\end{figure}

In contrast to HAT-P1, it almost certainly 
{\it will} be possible to put a lower bound on the 
value of $Q_p$ if a companion planet is detected and a reliable measurement
of the eccentricity of WASP-1b is made (as well as reliable estimates
for the companion's mass, 
semimajor axis and eccentricity). 
This is because $\tau_{circ}$ is only 14 Myr for $Q_p=10^5$, so that
$Q_p$ would have to be greater than $2.4\times 10^6$ for the system
to be substantially far from the equilibrium phase.
Figure~\ref{QpecWASP} 
is similar to Figure~\ref{Qpec}, however note the increase in $Q_p$
for the case $e_c$(now)=0.5, $a_c=0.3$AU. This is consistent
with the fact that $\tau_{circ}$ is relatively short, and since $\tau_c^{true}$
(equation~\rn{tautrue})
is proportional to $\tau_{circ}$, close companions will also circularize rather
quickly unless $Q_p$ is large.

\section{Circularizing the orbits of planets in the habitable zone}

\begin{figure}
\includegraphics[width=150mm]{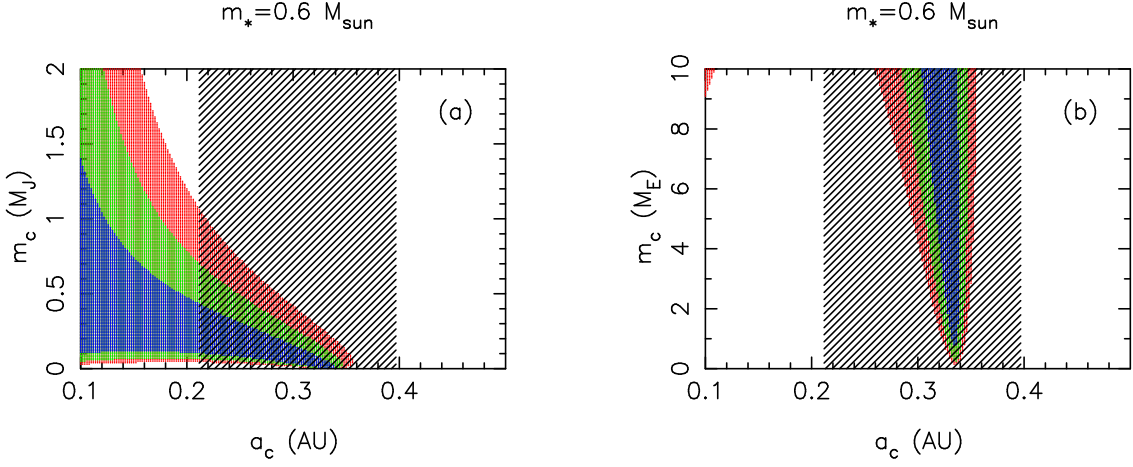}
\caption{Circularized planets in the habitable zone.
If the condition $t<t_{age}$ is met, a coloured dot is 
plotted according to the scheme: red for $t_{age}=6$ Gyr,
green for for $t_{age}=3$ Gyr, and blue for for $t_{age}=1$ Gyr.
A $Q$-value of $10^5$ and a radius of $1.3R_J$ are 
assumed for the short-period planet.
The hatched region corresponds to the habitable zone of the star.
For these systems, $a_p=0.04$, $m_p=0.6M_J$, 
$e_c^*=0.3$ and $m_*=0.6M_\odot$. Panel (b) shows detail
of panel (a) for low-mass planets ($M_E$ is the mass of the Earth).
}
\label{habitable1}
\end{figure}

\begin{figure}
\includegraphics[width=150mm]{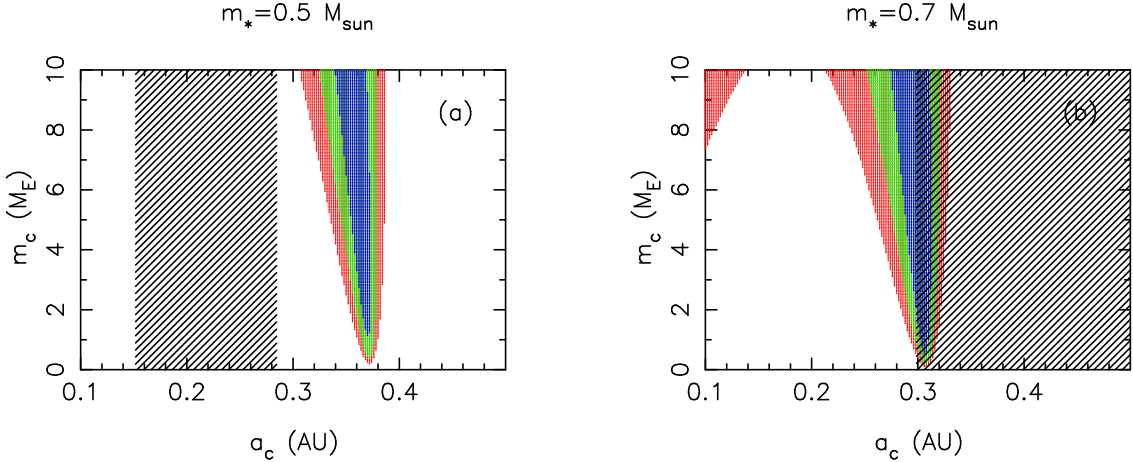}
\caption{Same parameters as in Figure~\ref{habitable1} but with
(a) $m_*=0.5M_\odot$ and (b) $m_*=0.7M_\odot$.}
\label{habitable2}
\end{figure}

Motivated by the work of \citet{r3}, \citet{m6} and \citet{f1} who find
Earth-mass planets forming externally to a migrating gas giant
after it scatters planetesimals outward past its own orbit,
we use the analysis presented here to identify two-planet configurations
for which the outer planet resides in the habitable zone of the system,
and for which the circularization timescale of that planet would
be longer than the age of the system without the assistance of a short-period
companion. Using equation \rn{ecaccurate} with $e_c=0.02$ (it is singular
at $e_c=0$), one can determine which values of $m_c/m_p$ and
$a_c/a_p$ produce $t<t_{age}$, where $t_{age}$ is the age of the star. 
Figure~\ref{habitable1} 
shows an example of this
for the case $a_p=0.04$, $m_p=0.6M_J$, $e_c^*=0.3$ and $m_*=0.6M_\odot$.
These parameters were chosen because the set of companion masses and 
semimajor axes for which the companion has already circularized 
contains a subset of Earth-mass planets which
lie inside the habitable zone of the star, defined to be such that the
semimajor axis lies in the range
\be
0.8\left(L_*/L_\odot\right)^{1/2}<a_c/AU<
1.5\left(L_*/L_\odot\right)^{1/2},
\ee
where $L_*$ is the luminosity of the star
and we have followed the definition given in \citet{s3}
which gives an empirical relationship between the luminosity
and the mass of a star. The habitable zone also contains a substantial
fraction of planets with masses up to $1M_J$; these systems are
also of interest because they can harbour moons which could
be habitable.
Figure~\ref{habitable2} 
shows the effect of decreasing and increasing
the stellar mass. Increasing $e_c^*$ pushes the minimum of the
circularized region upwards; increasing the mass of the short-period
planet pushes this region to the right as does
increasing $a_p$.

Currently, there are no known isolated short-period pairs of planets to
which this theory can be applied.
However, low-mass stars are now being targeted more frequently
so it seems likely such systems will be discovered in the near future.

\section{Summary and Discussion}\label{discussion}

\subsection{Summary}

The ideas and results presented in this paper are summarized as follows:

\ben
\item
If a planet is inflated because it is tidally heated, its radius is related
to the orbital eccentricity by equation~\rn{ept}. The biggest uncertainties
are the $Q$-value of the planet, $Q_p$, and the
heating needed to maintain the planet's size, ${\cal L}_p$. The latter
depends on the structural model used, and in particular, where inside the
planet the energy is dissipated. 
\item
Conditions for libration and circulation of $\eta=\varpi_p-\varpi_c$
including relativistic effects
are given by \rn{lib0r} and \rn{libpir} for the case that
$e_p\ll e_c$ and $m_p\ll m_*$ (there are no conditions on $m_c$), 
with the time dependence
of $e_p$, $e_c$ and $\eta$ given by
\rn{ep3}, \rn{ep4}, \rn{ect}, \rn{eta2}, \rn{eta3}, and \rn{etacirc},
with $e_p^{(eq)}$ given by \rn{equilGR}.
\item
The system evolves through three stages when tidal dissipation
is present (Figure~\ref{eta-GR-tides}): (1) circulation of 
the angle between the lines of apsides of the two planets accompanied
by the slow oscillation of the eccentricities at {\it constant amplitude},
together with decline
of the {\it mean} value of the inner planet's eccentricity until it reaches a (quasi)-fixed
value (equation \rn{edecline}): this occurs on the circularization timescale; 
(2) libration of the angle between the lines of apsides
accompanied by the slow oscillation of the eccentricities with reducing 
amplitude but maintenance of the mean value of the inner eccentricity
(equations \rn{eplib} and \rn{etalib} with the mean value $e_p^{(eq)}$
given by \rn{equilGR}):
this occurs on {\it twice} the circularization timescale; and (3) either
(a): a slow non-oscillatory decline in both eccentricities to zero or
(b): approximate maintenance of the inner eccentricity at the constant value
$e_p=e_p^*$ for some (often extremely long)
period of time followed by its {\it increase} to some maximum value
followed by an often fast decline of both eccentricities to zero. 
Both are described by \rn{equilGR} together with \rn{ecaccurate}.
Whether or not the final phase occurs via route (a) or (b), it proceeds 
on a timescale sometimes several orders of magnitude longer than
the tidal circularization timescale of the inner planet, $\tau_{circ}$.
\item
As a consequence of the existence of route (b) above,
very low-mass companions are sometimes capable of sustaining the eccentricity
needed to inflate short-period gas giants (eg.\ Figures~\ref{eup2} and \ref{eup3}).
For this to occur it is at least necessary that the companion
eccentricities are sufficiently high; the lower the companion mass
the higher the initial eccentricity needed.
The timescale on which the eccentricity is sustained increases
with {\it decreasing} companion mass, however, the value of the 
sustained eccentricity decreases with decreasing mass.
For a given system age, more solutions exist for higher values of $Q_p$.
This scenario is supported by the work of \citet{f1} and \citet{m6}
(see discussion below).
\item
As well as the very low-mass solutions produced by route (b),
route (a) produces
solutions for which the eccentricity and hence
radius of HD 209458b is sustained by a companion which produces
a stellar reflex velocity below the stellar jitter estimate 
(Figure~\ref{HD209458b}). More solutions
exist for higher $Q$-values of HD 209458b.
\item
Companion planet solutions are presented for HAT-P1 and WASP-1
for a wide range of companion parameters.
\item
If companions to HAT-P1b and/or WASP-1b (or other systems like these)
are discovered and reliable estimates of the system parameters can
be made (and there is no sign of any other companions), 
it should be possible to put lower bounds on the $Q$-value
of the short-period planet in the system (Figures~\ref{Qpec} and
\ref{QpecWASP}). Moderately inclined orbits have little effect
on the long-term evolution.
\item
Long after the protoplanetary disk has disappeared,
it is possible to circularize the orbits of some planets whose tidal
circularization timescales are longer than the age of the system
if they have short-period companions whose own values
of $\tau_{circ}$ are short. The circularization timescale of a
companion is given by equation~\rn{tautrue}, and depends
on the structure of the short-period planet.
\item
In the light of work by \citet{m6} and \citet{f1},
systems with hot Jupiters orbiting 
stars with masses around $0.5-0.7M_\odot$ are good candidates
for finding {\it circular} Earth-mass planets in the habitable zone,
as long as there is no third planet nearby.
\een

\subsection{Discussion}
How much the early evolution of short-period two-planet 
systems proceeds according
to the simple picture described here depends on how planets
come to be in their current positions. If a gaseous disk pilots them
in  \citep{l3} it will influence this phase at some level, especially in a case like
WASP-1b for which $\tau_{circ}$ is as low as 14 Myr (a time comparable
to the disk lifetime). 
If, instead, a planet is {\it scattered} in to its current position \citep{r2}
it is likely to arrive with significant eccentricity which must subsequently be
damped. Recall that the analysis here assumes $e_p\ll e_c$, although
relaxing this condition makes little difference to the overall picture.

In addition, conditions at the time of arrival will be very different to
those in a mature system. For example, a planet's radius 
will be significantly larger, thereby increasing the strength
of the tidal interaction with the star (see Fig.\ 1 in 
Bodenheimer, Lin \& Mardling 2001 for the time-dependence of the planetary
radius for several models). The stellar spin rate will 
be significantly higher, and in some cases be high
enough to force an {\it increase} in the planet's eccentricity.
The latter may account for the range of eccentricities observed in the
``borderline planets'' with periods in the range 7-21 days \citep{d2}.

It seems likely that Nature uses a hybrid of
migration and scattering, thereby producing a rich variety
of possible configurations \citep{m5,f1,m6,n1,z1}. 
Of particular interest here is the situation where planetesimals are 
shepherded in by a migrating gas giant, these going on either to accumulate
to form a hot Earth internal to the gas giant's orbit, 
or to be scattered out before accumulating
in an external orbit \citep{m6,f1}. In the former case the hot Earth
is often subsequently scattered out, so that both scenarios
produce eccentric low-mass objects with periods a few times
that of the massive planet. The final eccentricities of these Earth-mass planets
tend to be fairly low due to gas drag and dynamical friction with
nearby planetesimals, however the simulations 
of the above authors sampled only a small
range of initial conditions and it may be that substantially higher
eccentricities can result from this process. The maximum eccentricity
found by \citet{m6} was 0.46 for a planet with mass $0.32M_\oplus$ 
at $a_c=1.74$ AU outside a Jupiter-mass giant at $a_p=0.14$ AU
(simulation JD-3 in their Table 4). Higher eccentricities may
result when low-mass planets are scattered in via encounters
with bodies further out in the disk.
Note that the semimajor axis ratio in the above example was 12.4, about twice that
of the Earth-mass ``solutions'' for HD 209458 (Figure~\ref{HD209458d}),
while others were substantially lower.

While extremely low-mass solutions are relatively 
improbable, Figure~\ref{HD209458d} indicates that there are many
low-mass solutions for HD 209458 which are consistent with the findings of
\citet{f1} and \citet{m6}, especially for such $Q$-values ($3\times 10^5$).
While the findings of \citet{b3} and \citet{b2} suggest that
only low $Q$-values are consistent with the observed eccentricity
(if indeed it is non-zero), it should be remembered that these 
authors distributed the tidal energy homogeneously 
throughout the planet's envelope and implicitly assumed that the 
energy was dissipated locally. If, in reality, the energy is dissipated
near the surface (but not so close to the surface that there is
essentially no structural change \citep{w2}), then less tidal
energy may be needed to produce the observed radius for 
a given $Q$-value. This allows for higher $Q$-values to be 
associated with a given value of $e_p$ so that solutions
such as those shown in Figure~\ref{HD209458d} may, in fact, be
viable. {\it If so, inflated planets are good targets in the search
for terrestrial planets.}

Of course, while very low mass solutions are of extreme interest
(and completely unexpected!),
there are still many higher companion mass solutions for
HD 209458b, especially for 
$Q$-values above $10^5$. While WASP-1b and 
HAT-P1b remain unconstrained (except for the latter's eccentricity),
there exist almost unlimited potential solutions all of which offer
the opportunity to put lower bounds on their $Q$-values if 
companion planets are discovered. Moreover, if companions
are identified and their orbital parameters accurately
measured, the analysis presented here can be used to constrain their 
history, in particular, their ``initial'' eccentricity.

Conclusions such as these are possible as long as there is no third planet nearby,
and as long as one is confident that the circularization timescale (or
more precisely, {\it three times} the circularization timescale) is shorter
than the age of the system. Another possible departure from the assumptions
made in this analysis is non-coplanarity. However, unless 
the relative inclination of the two planets is substantial, say, greater,
than around $40^o$, non-coplanarity makes very little difference
to the predicted outcome, as has been verified by integration
of the full secular equations (Figure~\ref{relinc}).

Remarkably, at this stage there are no known two-planet systems
for which the innermost planet has a circularization timescale
less than a Gyr, and thus it is currently not possible
to compare our two-planet model with
a real system. In the meantime, 
equations \rn{equilGR} 
and \rn{ecexact} provide clear guidance on how such systems
depend on the system parameters.

\section*{Acknowledgments}

The author wishes to thank Doug Lin for valuable discussions and support,
and the referee Alexandre Correia who studied the paper thoroughly
and made very valuable suggestions.

\label{lastpage}

\end{document}